\documentclass[11pt,onecolumn,twoside]{IEEEtran}

\usepackage{cite}      
\usepackage{graphicx}  
\usepackage{amsmath}   
\usepackage{amssymb}   

\newtheorem{theorem}{Theorem}
\newtheorem{definition}{Definition} 

\begin{document}
\title{Quantization of Prior Probabilities \\for Hypothesis Testing}
\author{Kush~R.~Varshney and Lav~R.~Varshney%
\thanks{K.~R.~Varshney and L.~R.~Varshney are with the Laboratory for Information and Decision Systems, Massachusetts Institute of Technology, Cambridge, MA 02139 USA (e-mail: krv@mit.edu; lrv@mit.edu).}%
\thanks{Part of the material in this paper was presented at the 2008 IEEE International Conference on Acoustics, Speech, and Signal Processing \cite{VarshneyV2008}.}%
\thanks{This work was supported in part by NSF Graduate Research Fellowships to K.~R.~Varshney and L.~R.~Varshney.}}
\maketitle

\begin{abstract}
Bayesian hypothesis testing is investigated when the prior probabilities of the hypotheses, taken as a random vector, are quantized.  Nearest neighbor and centroid conditions are derived using mean Bayes risk error as a distortion measure for quantization.  A high-resolution approximation to the distortion-rate function is also obtained.  Human decision making in segregated populations is studied assuming Bayesian hypothesis testing with quantized priors.  
\end{abstract}


\section{Introduction}
\label{sec:intro}

\IEEEPARstart{C}{onsider} a hypothesis testing scenario in which an object is to be observed to determine which one of $M$ states, $\{h_0,\ldots,h_{M-1}\}$, it is in.  The object has prior probability $p_m$ of being in state $h_m$, i.e.~$p_m = \Pr[H=h_m]$, and prior probability vector $\mbox{\boldmath$p$} = \begin{bmatrix} p_0 & \cdots & p_{M-1} \end{bmatrix}^T$, with $\sum_{m=0}^{M-1} p_m = 1$, which is known to the decision maker.  $M$-ary hypothesis testing with known prior probabilities calls for the Bayesian formulation to the problem, for which the optimal decision rule minimizes Bayes risk \cite{WillskyWS2003}.  

Now consider the situation when there is a population of objects, each with its own prior probability vector drawn from the distribution $f_{\textsc{\mbox{\boldmath$P$}}}(\mbox{\boldmath$p$})$ supported on the $(M-1)$-dimensional probability simplex.  If the prior probability vector of each object were known perfectly to the decision maker before observation and hypothesis testing, then the scenario would be no different than that of standard Bayesian hypothesis testing.  However, we consider the case in which the decision maker is constrained and can only work with at most $K$ different prior probability vectors.  Such a constraint is motivated by scenarios where the decision maker has finite memory or limited information processing resources.  Hence, when there are more than $K$ objects in the population, the decision maker must first map the true prior probability vector of the object being observed to one of the $K$ available vectors and then proceed to perform the optimal Bayesian hypothesis test, treating that vector as the prior probabilities of the object.  

Although not the only such constrained scenario, one example is that of human decision making.  One particular setting is a referee deciding whether a player has committed a foul using his or her noisy observation as well as prior experience.  Players commit fouls at different rates; some players are dirtier or more aggressive than others.  It is this rate which is the prior probability for the `foul committed' state.  Hence, over the population of players, there is a distribution of prior probabilities.  If the referee tunes the prior probability to the particular player on whose action the decision is to be made, decision-making performance is improved.  

Human decision makers, however, are limited in their information processing capacity and can only carry around seven, plus or minus two, categories without getting confused \cite{Miller1956}.  Consequently, the referee is limited and categorizes players into a small number of dirtiness levels, with associated representative prior probabilities, exactly the scenario described above.  

In this paper, the design of the mapping from prior probability vectors in the population to one of $K$ representative probability vectors is approached as a quantization problem.  Mean Bayes risk error (MBRE) is defined as a fidelity criterion for the quantization of $f_{\textsc{\mbox{\boldmath$P$}}}(\mbox{\boldmath$p$})$ and conditions are derived for a minimum MBRE quantizer.  Some examples of MBRE-optimal quantizers are given along with their performance in the low-rate quantization regime.  Distortion-rate functions are given for the high-rate quantization regime.  Certain human decision-making tasks, as mentioned above, may be modeled by quantized prior hypothesis testing due to certain suboptimalities in human information processing.  Human decision making is analyzed in detail for segregated populations, revealing a mathematical model of social discrimination.  

Previous work that combines detection and quantization looks at the quantization of observed data, not prior probabilities, and also only approximates the Bayes risk function instead of working with it directly, e.g.~\cite{Kassam1977,PoorT1977,GuptaH2003} and references cited in \cite{GuptaH2003}.  In such work, there is a communication constraint between the sensor and the decision maker, but the decision maker has unconstrained processing capability.  Our work deals with the opposite case, where there is no communication constraint between the sensor and the decision maker, however the decision maker is constrained.  

A brief look at imperfect priors appears in \cite[Sec.~2.E]{Hildreth1963}, but optimal quantization is not considered.  In \cite{Kihlstrom1971,GillilandH1978}, it is shown that small deviations from the true prior yield small deviations in the Bayes risk.  We are not aware of any previous work that has looked at quantization, clustering, or categorization of prior probabilities.  

In the remainder of the paper, we focus on binary hypothesis testing, $M=2$.  Section~\ref{sec:bre} defines the Bayes risk error distortion and gives some of its properties.  Section~\ref{sec:lowrate} discusses low-rate quantization and Section~\ref{sec:highrate} discusses high-rate quantization.  Some examples with a Gaussian measurement model are given in Section~\ref{sec:example}.  Section~\ref{sec:impl} considers the implications on human decision making and Section~\ref{sec:conc} provides a summary and directions for future work.  

\section{Bayes Risk Error}
\label{sec:bre}

In the binary Bayesian hypothesis testing problem for a given object, there are two hypotheses $h_0$ and $h_1$ with prior probabilities $p_0 = \Pr[H=h_0]$ and $p_1 = \Pr[H=h_1] = 1 - p_0$, a noisy observation $Y$, and likelihoods $f_{Y|H}(y|h_0)$ and $f_{Y|H}(y|h_1)$.  Note that we consider a one-shot measurement $Y$, rather than a set of independent, noisy measurements.  A function $\hat{h}(y)$ is designed that uniquely maps every possible $y$ to either $h_0$ or $h_1$ in such a way that the function is optimal with respect to Bayes risk $J = E[ c(H_i,H_j)]$, an expectation over the non-negative cost function $c(h_i,h_j)$.  This gives the following specification for $\hat{h}(y)$:
\begin{equation}
\label{eq:argmin}
	\hat{h}(\cdot) = \arg\min_{f(\cdot)} E[c(H,f(Y))],
\end{equation}
where the expectation is over both $H$ and $Y$.  It may be shown that the optimal decision rule $\hat{h}(y)$ is the likelihood ratio test:
\begin{equation}
\label{eq:lrt}
	\frac{f_{Y|H}(y|h_1)}{f_{Y|H}(y|h_0)} \mathop{\lesseqgtr}^{\hat{h}(y)=h_1}_{\hat{h}(y)=h_0} \frac{p_0(c_{10}-c_{00})}{(1-p_0)(c_{01}-c_{11})},
\end{equation}
where $c_{ij} = c(h_i,h_j)$.  

There are two types of errors, with the following probabilities:
\begin{align*}
	p_E^{\text{I}} &= \Pr[\hat{h}(Y)=h_1 | H=h_0], \\
	p_E^{\text{II}} &= \Pr[\hat{h}(Y)=h_0 | H=h_1].
\end{align*}
Bayes risk may be expressed in terms of those error probabilities as:
\begin{equation}
\label{eq:bayesriskerrorprob}
	J = (c_{10}-c_{00})p_0p_E^{\text{I}} + (c_{01}-c_{11})(1-p_0)p_E^{\text{II}} + c_{00}p_0 + c_{11}(1-p_0).
\end{equation}
It is often of interest to assign no cost to correct decisions, i.e.~$c_{00} = c_{11} = 0$, which we assume in the remainder of this paper.  In this case, the Bayes risk simplifies to:
\begin{equation}
\label{eq:bayesrisk_f}
	J(p_0) = c_{10}p_0p_E^{\text{I}}(p_0) + c_{01}(1-p_0)p_E^{\text{II}}(p_0).
\end{equation} 	
In \eqref{eq:bayesrisk_f}, the dependence of the Bayes risk and error probabilities on $p_0$ has been explicitly noted.  The error probabilities depend on $p_0$ through $\hat{h}(\cdot)$, given in \eqref{eq:lrt}.  The function $J(p_0)$ is zero at the points $p_0=0$ and $p_0=1$ and is positive-valued, strictly concave, and continuous in the interval $(0,1)$ \cite{Wijsman1970,WillskyWS2003,DeGroot2004}.  

In the case when the true prior probability is $p_0$, but $\hat{h}(y)$ is designed according to \eqref{eq:lrt} using some other value $a$ substituted for $p_0$, there is mismatch, and the mismatched Bayes risk is:
\begin{equation}
\label{eq:bayesrisk_mismatch}
	\tilde{J}(p_0,a) = c_{10}p_0p_E^{\text{I}}(a) + c_{01}(1-p_0)p_E^{\text{II}}(a).
\end{equation}
$\tilde{J}(p_0,a)$ is a linear function of $p_0$ with slope $(c_{10}p_E^{\text{I}}(a) - c_{01}p_E^{\text{II}}(a))$ and intercept $c_{01}p_E^{\text{II}}(a)$.  Note that $\tilde{J}(p_0,a)$ is tangent to $J(p_0)$ at $a$ and that $\tilde{J}(p_0,p_0) = J(p_0)$.

\begin{definition}
Let Bayes risk error $d(p_0,a)$ be the difference between the mismatched Bayes risk function $\tilde{J}(p_0,a)$ and the Bayes risk function $J(p_0)$:
\begin{align}
\label{eq:bre}
	d(p_0,a) &= \tilde{J}(p_0,a) - J(p_0) \nonumber\\ 
		&= c_{10}p_0p_E^{\text{I}}(a) + c_{01}(1-p_0)p_E^{\text{II}}(a) - c_{10}p_0p_E^{\text{I}}(p_0) - c_{01}(1-p_0)p_E^{\text{II}}(p_0).
\end{align}
\end{definition}

We now give properties of $d(p_0,a)$ as a function of $p_0$ and as a function of $a$.

\begin{theorem}
\label{thm:p0}
The Bayes risk error $d(p_0,a)$ is non-negative and only equal to zero when $p_0 = a$.  As a function of $p_0 \in (0,1)$, it is continuous and strictly convex for all $a$.
\end{theorem}

\begin{IEEEproof}
Since $J(p_0)$ is a continuous and strictly concave function, and lines $\tilde{J}(p_0,a)$ are tangent to $J(p_0)$, $\tilde{J}(p_0,a) \ge J(p_0)$ for all $p_0$ and $a$, with equality when $p_0 = a$.  Consequently, $d(p_0,a)$ is non-negative and only equal to zero when $p_0 = a$.  Moreover, $d(p_0,a)$ is continuous and strictly convex in $p_0 \in (0,1)$ for all $a$ because it is the difference of a continuous linear function and a continuous strictly concave function.  
\end{IEEEproof}

\begin{theorem}
\label{thm:a} 
For any deterministic likelihood ratio test $\hat{h}(\cdot)$, as a function of $a \in (0,1)$ for all $p_0$, the Bayes risk error $d(p_0,a)$ has exactly one stationary point, which is a minimum.  
\end{theorem}

\begin{IEEEproof}
Consider the parameterized curve $(p_E^{\text{I}},p_E^{\text{II}})$ traced out as $a$ is varied; this is a flipped version of the receiver operating characteristic (ROC).  The flipped ROC is a strictly convex function for deterministic likelihood ratio tests.  At its endpoints, it takes values $(p_E^{\text{I}} = 0, p_E^{\text{II}} = 1)$ when $a=1$ and $(p_E^{\text{I}} = 1, p_E^{\text{II}} = 0)$ when $a=0$ \cite{WillskyWS2003}, and therefore has average slope $-1$.  By the mean value theorem and strict convexity, there exists a unique point on the flipped ROC at which $\frac{dp_E^{\text{II}}}{dp_E^{\text{I}}} = -1$.  To the left of that point: $-\infty < \frac{dp_E^{\text{II}}}{dp_E^{\text{I}}} < -1$, and to the right of that point: $-1 < \frac{dp_E^{\text{II}}}{dp_E^{\text{I}}} < 0$.  

For deterministic likelihood ratio tests, $\beta\frac{dp_E^{\text{I}}(a)}{da} < 0$ and $\gamma\frac{dp_E^{\text{II}}(a)}{da} > 0$ for all $a \in (0,1)$ and positive constants $\beta$ and $\gamma$ \cite{WillskyWS2003}.  Therefore, if $\frac{\gamma dp_E^{\text{II}}}{\beta dp_E^{\text{I}}} < -1$, i.e.~$\frac{\gamma dp_E^{\text{II}}}{da}\frac{da}{\beta dp_E^{\text{I}}} < -1$, then $\gamma\frac{dp_E^{\text{II}}}{da} > -\beta\frac{dp_E^{\text{I}}}{da}$ and $\beta\frac{dp_E^{\text{I}}}{da} + \gamma\frac{dp_E^{\text{II}}}{da} > 0$.  In the same manner, if $\frac{\gamma dp_E^{\text{II}}}{\beta dp_E^{\text{I}}} > -1$, then $\beta\frac{dp_E^{\text{I}}}{da} + \gamma\frac{dp_E^{\text{II}}}{da} < 0$.  

Combining the above, we find that the function $\beta p_E^{\text{I}}(a) + \gamma p_E^{\text{II}}(a)$ has exactly one stationary point in $(0,1)$, which occurs when the slope of the flipped ROC is $-\frac{\beta}{\gamma}$.  Denote this stationary point as $a_s$.  For $0 < a < a_s$, $-1 < \frac{dp_E^{\text{II}}}{dp_E^{\text{I}}} < 0$ and the slope of $\beta p_E^{\text{I}}(a) + \gamma p_E^{\text{II}}(a)$ is negative; for $a_s < a < 1$, $-\infty < \frac{dp_E^{\text{II}}}{dp_E^{\text{I}}} < -1$ and the slope of $\beta p_E^{\text{I}}(a) + \gamma p_E^{\text{II}}(a)$ is positive.  Therefore, $a_s$ is a minimum.  

As a function of $a$, the Bayes risk error is of the form $\beta p_E^{\text{I}}(a) + \gamma p_E^{\text{II}}(a) + C$.  Hence, it also has exactly one stationary point $a_s$, which is a minimum.  
\end{IEEEproof}

As seen in Section~\ref{sec:lowrate}, the above properties of $d(p_0,a)$ are useful to establish that the Lloyd-Max conditions are not only necessary, but also sufficient for quantizer local optimality.

The third derivative of $d(p_0,a)$ with respect to $p_0$ is:
\begin{equation}
\label{eq:thirdderiv}
	-c_{10} p_0 \tfrac{d^3 p_E^{\text{I}}(p_0)}{dp_0^3} - 3 c_{10} \tfrac{d^2 p_E^{\text{I}}(p_0)}{dp_0^2} - c_{01}(1-p_0)\tfrac{d^3 p_E^{\text{II}}(p_0)}{dp_0^3} + 3 c_{01}\tfrac{d^2 p_E^{\text{II}}(p_0)}{dp_0^2},
\end{equation}
when the constituent derivatives exist.  As seen in Section \ref{sec:highrate}, when the third derivative exists and is continuous, $d(p_0,a)$ is locally quadratic, which is useful to develop high-rate quantization theory for Bayes risk error fidelity \cite{LiCG1999}.  

\section{Low-Rate Quantization}
\label{sec:lowrate}

The conditions necessary for the optimality of a quantizer for $f_{P_0}(p_0)$ under Bayes risk error distortion are now derived.  A $K$-point quantizer partitions the interval $[0,1]$ into $K$ regions $\mathcal{R}_1$, $\mathcal{R}_2$, $\mathcal{R}_3$, \ldots, $\mathcal{R}_K$.  For each of these quantization regions $\mathcal{R}_k$, there is a representation point $a_k$ to which elements are mapped.  For regular quantizers, the regions are  subintervals $\mathcal{R}_1 = [0,b_1]$, $\mathcal{R}_2 = (b_1,b_2]$, $\mathcal{R}_3 = (b_2,b_3]$, \ldots, $\mathcal{R}_K = (b_{K-1},1]$ and the representation points $a_k$ are in $\mathcal{R}_k$.\footnote{Due to the strict convexity of $d(p_0,a)$ in $p_0$ for all $a$ shown in Theorem~\ref{thm:p0}, quantizers that satisfy the necessary conditions for MBRE optimality are regular, see \cite[Lemma 6.2.1]{GershoG1992}.  Therefore, only regular quantizers are considered.}  A quantizer can be viewed as a nonlinear function $v_K(\cdot)$ such that $v_K(p_0) = a_k$ for $p_0 \in \mathcal{R}_k$.  For a given $K$, we would like to find the quantizer that minimizes the MBRE:
\begin{equation}
\label{eq:MBRE}
	D = E[d(P_0,v_K(P_0))] = \int d(p_0,v_K(p_0)) f_{P_0}(p_0)dp_0.
\end{equation}
There is no closed-form solution, but an optimal quantizer must satisfy the nearest neighbor condition, the centroid condition, and the zero probability of boundary condition \cite{GershoG1992}.  The nearest neighbor and centroid conditions are developed for MBRE in the following subsections.  When $f_{P_0}(p_0)$ is absolutely continuous, the zero probability of boundary condition is always satisfied.  

\subsection{Nearest Neighbor Condition}
\label{sec:lowrate:nearest}

\begin{figure}
	\begin{center}
		\includegraphics[width=0.48\textwidth]{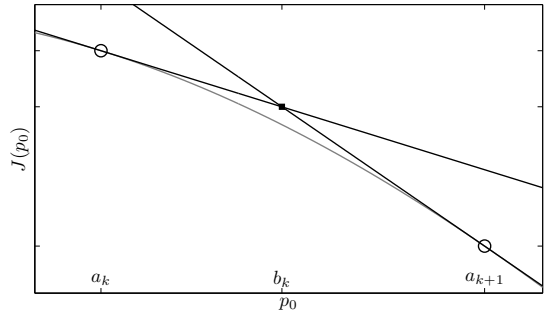} 
	\end{center}
	\caption{The intersection of the lines $\tilde{J}(p_0,a_k)$, tangent to $J(p_0)$ at $a_k$, and $\tilde{J}(p_0,a_{k+1})$, tangent to $J(p_0)$ at $a_{k+1}$, is the optimal interval boundary.}
\label{fig:intersection}
\end{figure}

With the representation points $\{a_k\}$ fixed, an expression for the interval boundaries $\{b_k\}$ is derived.  Given any $p_0 \in [a_k,a_{k+1}]$, if $\tilde{J}(p_0,a_k) < \tilde{J}(p_0,a_{k+1})$ then Bayes risk error is minimized if $p_0$ is represented by $a_k$, and if $\tilde{J}(p_0,a_k) > \tilde{J}(p_0,a_{k+1})$ then Bayes risk error is minimized if $p_0$ is represented by $a_{k+1}$.  The boundary point $b_k \in [a_k,a_{k+1}]$ is the abscissa of the point at which the lines $\tilde{J}(p_0,a_k)$ and $\tilde{J}(p_0,a_{k+1})$ intersect.  The idea is illustrated graphically in Fig.~\ref{fig:intersection}.

By manipulating the slopes and intercepts of $\tilde{J}(p_0,a_k)$ and $\tilde{J}(p_0,a_{k+1})$, the point of intersection is found to be:
\begin{equation}
\label{eq:bj}
	b_k = \frac{c_{01}\left(p_E^{\text{II}}(a_{k+1}) - p_E^{\text{II}}(a_k)\right)}{c_{01}\left(p_E^{\text{II}}(a_{k+1}) - p_E^{\text{II}}(a_k)\right) - c_{10}\left(p_E^{\text{I}}(a_{k+1}) - p_E^{\text{I}}(a_k)\right)}.
\end{equation}

\subsection{Centroid Condition}
\label{sec:lowrate:centroid}

With the quantization regions fixed, the MBRE is to be minimized over the $\{a_k\}$.  Here, the MBRE is expressed as the sum of integrals over quantization regions:
\begin{equation}
\label{eq:MBRE_centroid}
	D = \sum_{k=1}^K \int_{\mathcal{R}_k}\left(\tilde{J}(p_0,a_k)-J(p_0)\right)f_{P_0}(p_0)dp_0.
\end{equation}
Because the regions are fixed, the minimization may be performed for each interval separately.  

Let us define $I^{\text{I}}_k = \int_{\mathcal{R}_k} p_0 f_{P_0}(p_0)dp_0$ and $I^{\text{II}}_k = \int_{\mathcal{R}_k}(1-p_0) f_{P_0}(p_0)dp_0$, which are conditional means.  Then:
\begin{equation}
\label{eq:minimization2}
	a_k = \arg\min_a \left\{c_{10}I^{\text{I}}_kp_E^{\text{I}}(a) + c_{01}I^{\text{II}}_kp_E^{\text{II}}(a)\right\}.
\end{equation}

Since $\beta p_E^{\text{I}}(a) + \gamma p_E^{\text{II}}(a)$ has exactly one stationary point, which is a minimum (cf. Theorem~\ref{thm:a}), equation~\eqref{eq:minimization2} is uniquely minimized by setting its derivative equal to zero.  Thus, $a_k$ is the solution to:
\begin{equation}
\label{eq:aj}
	c_{10}I^{\text{I}}_k\left.\tfrac{dp_E^{\text{I}}(a)}{da}\right|_{a_k} + c_{01}I^{\text{II}}_k\left.\tfrac{dp_E^{\text{II}}(a)}{da}\right|_{a_k} = 0.
\end{equation}
Commonly, differentiation of the two error probabilities is tractable; they are themselves integrals of the likelihood functions and the differentiation is with respect to some function of the limits of integration.  

\subsection{Lloyd-Max Algorithm}
\label{sec:lowrate:lm}

Alternating between the nearest neighbor and centroid conditions, the iterative Lloyd-Max algorithm can be applied to find minimum MBRE quantizers \cite{GershoG1992}.  The algorithm is widely used because of its simplicity, effectiveness, and convergence properties \cite{GrayN1998}.  

In \cite{Trushkin1982}, it is shown that the conditions necessary for optimality of the quantizer are also sufficient conditions for local optimality\footnote{By local optimality, it is meant that the $\{a_k\}$ and $\{b_k\}$ minimize the objective function \eqref{eq:MBRE} among feasible representation and boundary points near them.} if the following hold.  The first condition is that $f_{P_0}(p_0)$ must be positive and continuous in $(0,1)$.  The second condition is that $\int_0^1 d(p_0,a)f_{P_0}(p_0)dp_0$ must be finite for all $a$.  The first and second conditions are met by common distributions such as the beta distribution \cite{Fine2006}.  

The third condition is that the distortion function $d(p_0,a)$ must satisfy some properties.  It must be zero only for $p_0 = a$, continuous in $p_0$ for all $a$, and convex in $a$; the first two of these hold as discussed in Theorem~\ref{thm:p0}.  The third, convexity in $a$, does not hold for Bayes risk error in general, but the convexity of $d(p_0,a)$ in $a$ is only used by \cite{Trushkin1982} to show that a unique minimum exists.  As shown in Theorem~\ref{thm:a}, $d(p_0,a)$ has a unique stationary point that is a minimum.  Therefore, the analysis of \cite{Trushkin1982} applies to Bayes risk error distortion.  Thus, if $f_{P_0}(p_0)$ satisfies the first and second conditions, then the algorithm is guaranteed to converge to a local optimum.  The algorithm may be run many times with different initializations to find the global optimum.  

Further conditions on $d(p_0,a)$ and $f_{P_0}(p_0)$ are given in \cite{Trushkin1982} for there to be a unique locally optimal quantizer, i.e.~the global optimum.  If these further conditions for unique local optimality hold, then the algorithm is guaranteed to find the globally minimum MBRE quantizer.  

In many practical situations, the distribution $f_{P_0}(p_0)$ is not available, but data drawn from it is available.  The optimal design of quantizers from data is NP-hard \cite{GareyJW1982,DrineasFKVV1999}.  However, the Lloyd-Max algorithm and its close cousin $K$-means can be used on data with the Bayes risk error fidelity criterion.  In fact, as the size of the dataset increases, the sequence of quantizers designed from data converges to the quantizer designed from $f_{P_0}(p_0)$ \cite{GrayKL1980,SabinG1986}.  (Conditions on the distortion function given in \cite{SabinG1986} except convexity in $a$ are met by the Bayes risk error, but in a similar way to the sufficiency of the Lloyd-Max conditions, the unique minimum property of the Bayes risk error is enough.)  

\subsection{Monotonic Convergence in $K$}
\label{sec:lowrate:convergence}

Let $D^*(K) = \sum_{k=1}^K \int_{\mathcal{R}_k^*}d(p_0,a_k^*)f_{P_0}(p_0)dp_0$ denote the MBRE for an optimal $K$-point quantizer.  We show that $D^*(K)$ monotonically converges as $K$ increases.  The MBRE-optimal $K$-point quantizer is the solution to the following problem:
\begin{align}
\label{eq:generalmin}
	\text{minimize}\quad &\sum_{k=1}^K \int_{b_{k-1}}^{b_k} d(p_0,a_k)f_{P_0}(p_0)dp_0 \nonumber\\
	\text{such that}\quad &b_0 = 0 \nonumber\\
				&b_K = 1 \nonumber\\
				&b_{k-1} \le a_k, \; k = 1,\ldots, K \nonumber\\
				&a_k \le b_k, \; k = 1,\ldots, K. 
\end{align}
Let us add the additional constraint $b_{K-1} = 1$ to \eqref{eq:generalmin}, forcing $a_K = 1$ and degeneracy of the $K^\text{th}$ quantization region.  The optimization problem for the $K$-point quantizer \eqref{eq:generalmin} with the additional constraint is equivalent to the optimization problem for the ($K$$-$$1$)-point quantizer.  Thus, the ($K$$-$$1$)-point design problem and the $K$-point design problem have the same objective function, but the ($K$$-$$1$)-point problem has an additional constraint.  Therefore, $D^*(K-1) \ge D^*(K)$.  

Since $d(p_0,v_K(p_0)) \ge 0$, $D = E[d(P_0,v_K(P_0))] \ge 0$.  Since the sequence $D^*(K)$ is nonincreasing and bounded from below, it converges.  Mean Bayes risk error cannot get worse when more quantization levels are employed.  In typical settings, as in Section \ref{sec:example}, performance always improves with an increase in the number of quantization levels.  

\section{High-Rate Quantization}
\label{sec:highrate}

Let us apply high-rate quantization theory \cite{GrayN1998} to the study of minimum MBRE quantization.  The distortion function for the MBRE criterion has a positive second derivative in $p_0$ (due to strict convexity) and for many families of likelihood functions, it has a continuous third derivative, see \eqref{eq:thirdderiv}.  Thus, it is locally quadratic in the sense of Li \emph{et} \emph{al.} \cite{LiCG1999} and in a manner similar to many perceptual, non-difference distortion functions, the high-rate quantization theory is well-developed.  

At high rate, i.e.~$K$ large, if we let:
\begin{equation}
\label{eq:B}
	B(p_0) = -\tfrac{1}{2}c_{10} p_0 \tfrac{d^2 p_E^{\text{I}}(p_0)}{dp_0^2} - c_{10} \tfrac{d p_E^{\text{I}}(p_0)}{dp_0} - \tfrac{1}{2}c_{01}(1-p_0)\tfrac{d^2 p_E^{\text{II}}(p_0)}{dp_0^2} + c_{01}\tfrac{d p_E^{\text{II}}(p_0)}{dp_0},
\end{equation}
then $d(p_0,a_k)$ is approximated by the following second order Taylor expansion:
\begin{equation}
\label{eq:breapprox}
	d(p_0,a_k) \approx \left.B(p_0)\right|_{p_0=a_k}\left(p_0-a_k\right)^2, \quad p_0 \in \mathcal{R}_k.
\end{equation}
Assuming that $f_{P_0}(\cdot)$ is sufficiently smooth and substituting \eqref{eq:breapprox} into the objective of \eqref{eq:generalmin}, the MBRE is approximated by:
\begin{equation}
\label{eq:mbreapprox}
	D \approx \sum_{k=1}^K f_{P_0}(a_k) B(a_k) \int_{\mathcal{R}_k} \left(p_0-a_k \right)^2dp_0.
\end{equation}
The MBRE is greater than and approximately equal to the following lower bound, derived in \cite{LiCG1999} by relationships involving normalized moments of inertia of intervals $\mathcal{R}_k$ and by H\"{o}lder's inequality:
\begin{equation}
\label{eq:mbreapprox2}
	D_L = \tfrac{1}{12K^2}\int_0^1 B(p_0)f_{P_0}(p_0)\lambda(p_0)^{-2}dp_0,
\end{equation}
where the optimal quantizer point density is:
\begin{equation}
\label{eq:pointdensity}
	\lambda(p_0) = \frac{\left(B(p_0)f_{P_0}(p_0)\right)^{1/3}}{\int_0^1 \left(B(p_0)f_{P_0}(p_0)\right)^{1/3} dp_0}.
\end{equation}
Integrating a quantizer point density over an interval yields the fraction of the $\{a_k\}$ that are in that interval.  Substituting \eqref{eq:pointdensity} into \eqref{eq:mbreapprox2} yields:
\begin{equation}
\label{eq:mbreapprox3}
	D_L = \tfrac{1}{12K^2}\|B(p_0)f_{P_0}(p_0)\|_{1/3}.
\end{equation}

\section{Examples}
\label{sec:example}

As an example, let us consider the following scalar signal and measurement model:
\begin{equation}
\label{eq:scalarmeas}
	Y = s_m + W, \quad m \in \{0,1\},
\end{equation}
where $s_0=0$ and $s_1=\mu$ (a known, deterministic quantity), and $W$ is a zero-mean, Gaussian random variable with variance $\sigma^2$.  The likelihoods are:
\begin{align}
\label{eq:likelihoods}
	f_{Y|H}(y|h_0) &= \mathcal{N}(y;0,\sigma^2) = \tfrac{1}{\sigma\sqrt{2\pi}} e^{-y^2/2\sigma^2}, \nonumber \\
	f_{Y|H}(y|h_1) &= \mathcal{N}(y;\mu,\sigma^2) = \tfrac{1}{\sigma\sqrt{2\pi}} e^{-(y-\mu)^2/2\sigma^2}.
\end{align}
The two error probabilities are:
\begin{align}
\label{eq:errors_gauss}
	p_E^{\text{I}}(p_0) &= Q\left( \tfrac{\mu}{2\sigma} + \tfrac{\sigma}{\mu}\ln\left(\tfrac{c_{10}p_0}{c_{01}(1-p_0)}\right)\right), \nonumber \\
	p_E^{\text{II}}(p_0) &= Q\left( \tfrac{\mu}{2\sigma} - \tfrac{\sigma}{\mu}\ln\left(\tfrac{c_{10}p_0}{c_{01}(1-p_0)}\right)\right),
\end{align}
where:
\begin{displaymath}
	Q(\alpha) = \tfrac{1}{\sqrt{2\pi}} \int_\alpha^\infty e^{-x^2/2} dx.
\end{displaymath}

Finding the centroid condition, the derivatives of the error probabilities are:
\begin{align}
\label{eq:gauss_derivsI}
	\left.\tfrac{dp_E^{\text{I}}(p_0)}{dp_0}\right|_{a_k} &= -\tfrac{1}{\sqrt{2\pi}}\tfrac{\sigma}{\mu}\tfrac{1}{a_k(1-a_k)}e^{-\frac{1}{2}\left(\frac{\mu}{2\sigma} + \frac{\sigma}{\mu}\ln\left(\frac{c_{10}a_k}{c_{01}(1-a_k)}\right)\right)^2}, \\
\label{eq:gauss_derivsII}
	\left.\tfrac{dp_E^{\text{II}}(p_0)}{dp_0}\right|_{a_k} &= +\tfrac{1}{\sqrt{2\pi}}\tfrac{\sigma}{\mu}\tfrac{1}{a_k(1-a_k)}e^{-\frac{1}{2}\left(\frac{\mu}{2\sigma} - \frac{\sigma}{\mu}\ln\left(\frac{c_{10}a_k}{c_{01}(1-a_k)}\right)\right)^2}.
\end{align}
By substituting these derivatives into \eqref{eq:aj} and simplifying, the following expression is obtained for the representation points:
\begin{equation}
\label{eq:aj_gauss}
	a_k = \frac{I^{\text{I}}_k}{I^{\text{I}}_k+I^{\text{II}}_k}.
\end{equation}

For high-rate analysis, the second derivatives of the error probabilities are needed.  They are:
\begin{equation}
\label{eq:gauss_2derivsI}
	\tfrac{d^2p_E^{\text{I}}(p_0)}{dp_0^2} = -\tfrac{1}{\sqrt{8\pi}}\tfrac{\sigma}{\mu}\tfrac{1}{p_0^2(1-p_0)^2}e^{-\frac{1}{8\mu^2\sigma^2}\left(\mu^2 + 2\sigma^2\ln\left(\frac{c_{10}p_0}{c_{01}(1-p_0)}\right)\right)^2}\left[-3 + 4p_0 - \tfrac{2\sigma^2}{\mu^2}\ln\left(\tfrac{c_{10}p_0}{c_{01}(1-p_0)}\right)\right],
\end{equation}
and:
\begin{equation}
\label{eq:gauss_2derivsII}
	\tfrac{d^2p_E^{\text{II}}(p_0)}{dp_0^2} = +\tfrac{1}{\sqrt{8\pi}}\tfrac{\sigma}{\mu}\tfrac{1}{p_0^2(1-p_0)^2}e^{-\frac{1}{8\mu^2\sigma^2}\left(\mu^2 - 2\sigma^2\ln\left(\frac{c_{10}p_0}{c_{01}(1-p_0)}\right)\right)^2}\left[-1 + 4p_0 - \tfrac{2\sigma^2}{\mu^2}\ln\left(\tfrac{c_{10}p_0}{c_{01}(1-p_0)}\right)\right].
\end{equation}
By inspection, we note that the third derivatives are continuous.  Substituting the first derivatives \eqref{eq:gauss_derivsI}-\eqref{eq:gauss_derivsII} and second derivatives \eqref{eq:gauss_2derivsI}-\eqref{eq:gauss_2derivsII} into \eqref{eq:B}, an expression for $B(p_0)$ can be obtained.  

Examples with different distributions $f_{P_0}(p_0)$ are presented below.  All of the examples use scalar signals with additive Gaussian noise, $\mu = 1$, $\sigma = 1$ \eqref{eq:scalarmeas}.  As a point of reference, a comparison is made to quantizers designed under mean absolute error (MAE) \cite{Kassam1978}, i.e.~$d(p_0,a) = |p_0-a|$, an objective that does not account for hypothesis testing.\footnote{As shown by Kassam \cite{Kassam1978}, minimizing the MAE criterion also minimizes the absolute distance between the cumulative distribution function of the source and the induced cumulative distribution function of the quantized output.  Since the induced distribution from quantization is used as the population prior distribution for hypothesis testing, requiring this induced distribution to be close to the true unquantized distribution is reasonable.  If distance between probability distributions is to be minimized according to the Kullback-Leibler discrimination between the true and induced distributions (which is defined in terms of likelihood ratios), an application of Pinsker's inequality shows that a small absolute difference is requisite \cite{Topsoe2000}.  Although a reasonable criterion, MAE is suboptimal for hypothesis testing performance as seen in the examples.}  

In the high-rate comparisons, the optimal point density for MAE \cite{GrayG1977}:
\begin{displaymath}
	\lambda(p_0) = \frac{f_{P_0}(p_0)^{1/2}}{\int_0^1 f_{P_0}(p_0)^{1/2} dp_0}
\end{displaymath}
is substituted into the high-rate distortion approximation for the MBRE criterion \eqref{eq:mbreapprox2}.  Taking $R = \log_2(K)$, there is a constant gap between the rates using the MBRE point density and the MAE point density for all distortion values.  This difference is:
\begin{displaymath}
	R_{\text{MBRE}}(D_L) - R_{\text{MAE}}(D_L) = \frac{1}{2}\log_2\left(\frac{\|f_{P_0}(p_0)B(p_0)\|_{1/3}}{\|f_{P_0}(p_0)\|_{1/2}\int_0^1 B(p_0) dp_0}\right).
\end{displaymath}
The closer the ratio inside the logarithm is to one, the closer the MBRE- and MAE-optimal quantizers.

\subsection{Uniformly Distributed $P_0$}
\label{sec:example:uniform}

\begin{figure}
	\begin{center}
		\includegraphics[width=0.48\textwidth]{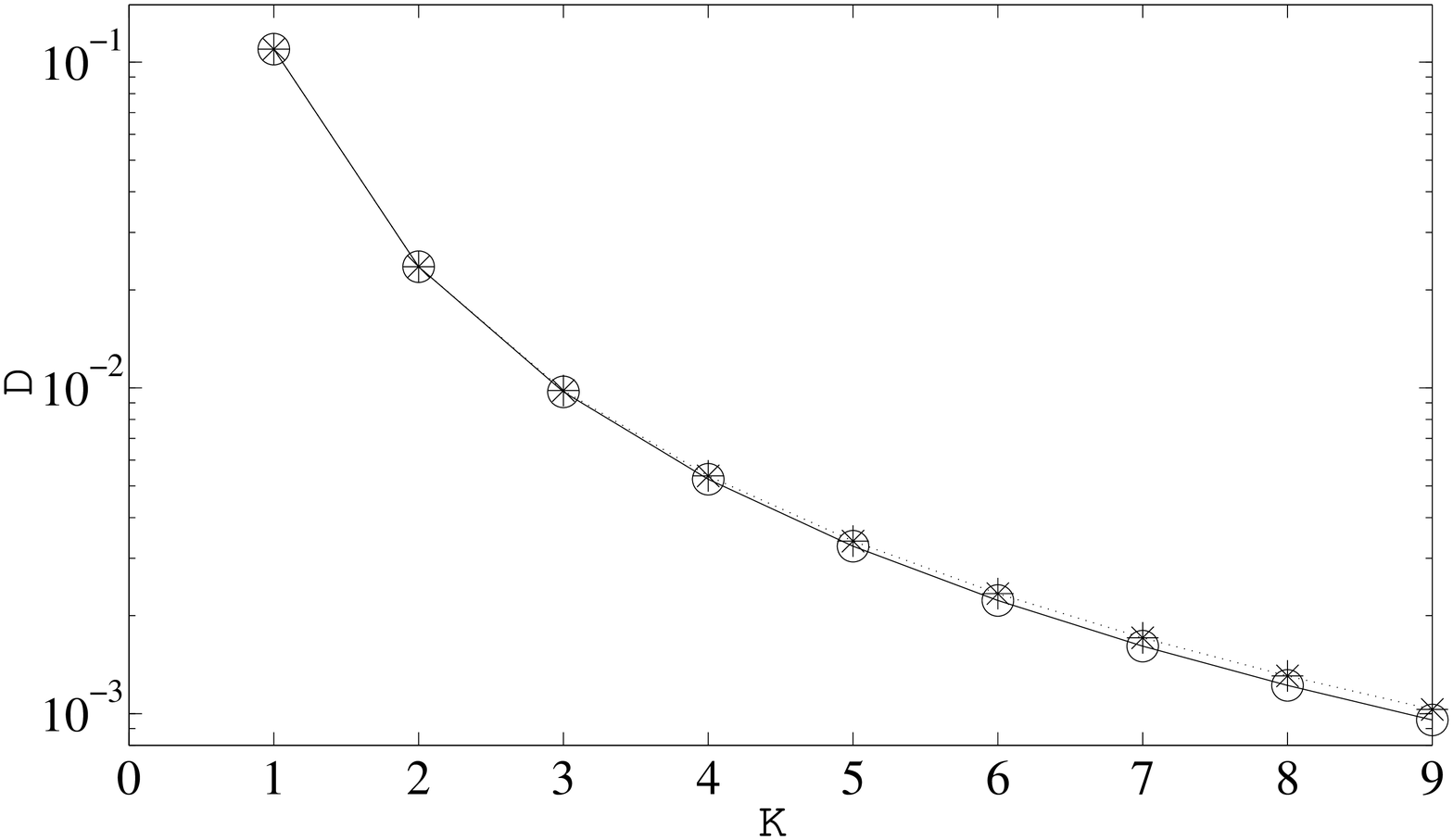}
	\end{center}
	\caption{MBRE for uniformly distributed $P_0$ and Bayes costs $c_{10} = c_{01} = 1$ plotted on a logarithmic scale as a function of the number of quantization levels $K$; the solid line with circle markers is the MBRE-optimal quantizer and the dotted line with asterisk markers is the MAE-optimal uniform quantizer.}
\label{fig:MBRE_uniform}
\end{figure}

\begin{figure}
	\begin{center}
		\includegraphics[width=0.48\textwidth]{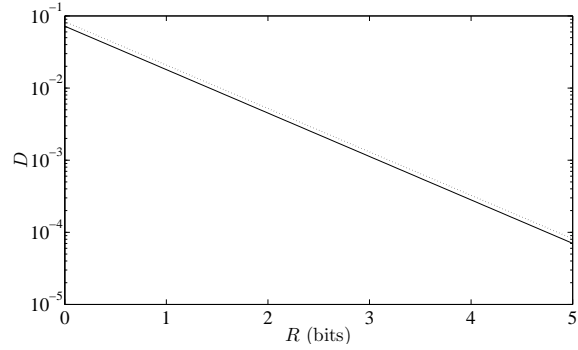}
	\end{center}
	\caption{High-rate approximation of distortion-rate function $D_L$ for uniformly distributed $P_0$ and Bayes costs $c_{10} = c_{01} = 1$; the solid line is the MBRE-optimal quantizer and the dotted line is the MAE-optimal uniform quantizer.}
\label{fig:rd_uniform}
\end{figure}

We first look at the setting in which all prior probabilities are equally likely.  The MBRE of the MBRE-optimal quantizer and a quantizer designed to minimize MAE with respect to $f_{P_0}(p_0)$ are plotted in Fig.~\ref{fig:MBRE_uniform}.  
(The optimal MAE quantizer for the uniform distribution is the uniform quantizer.)  The plot shows MBRE as a function of $K$; the solid line with circle markers is the MBRE-optimal quantizer and the dotted line with asterisk markers is the MAE-optimal quantizer.  $D_L$, the high-rate approximation to the distortion-rate function is plotted in Fig.~\ref{fig:rd_uniform}.  

\begin{figure}
	\begin{center}
		\begin{tabular}{cc}
			\includegraphics[width=0.22\textwidth]{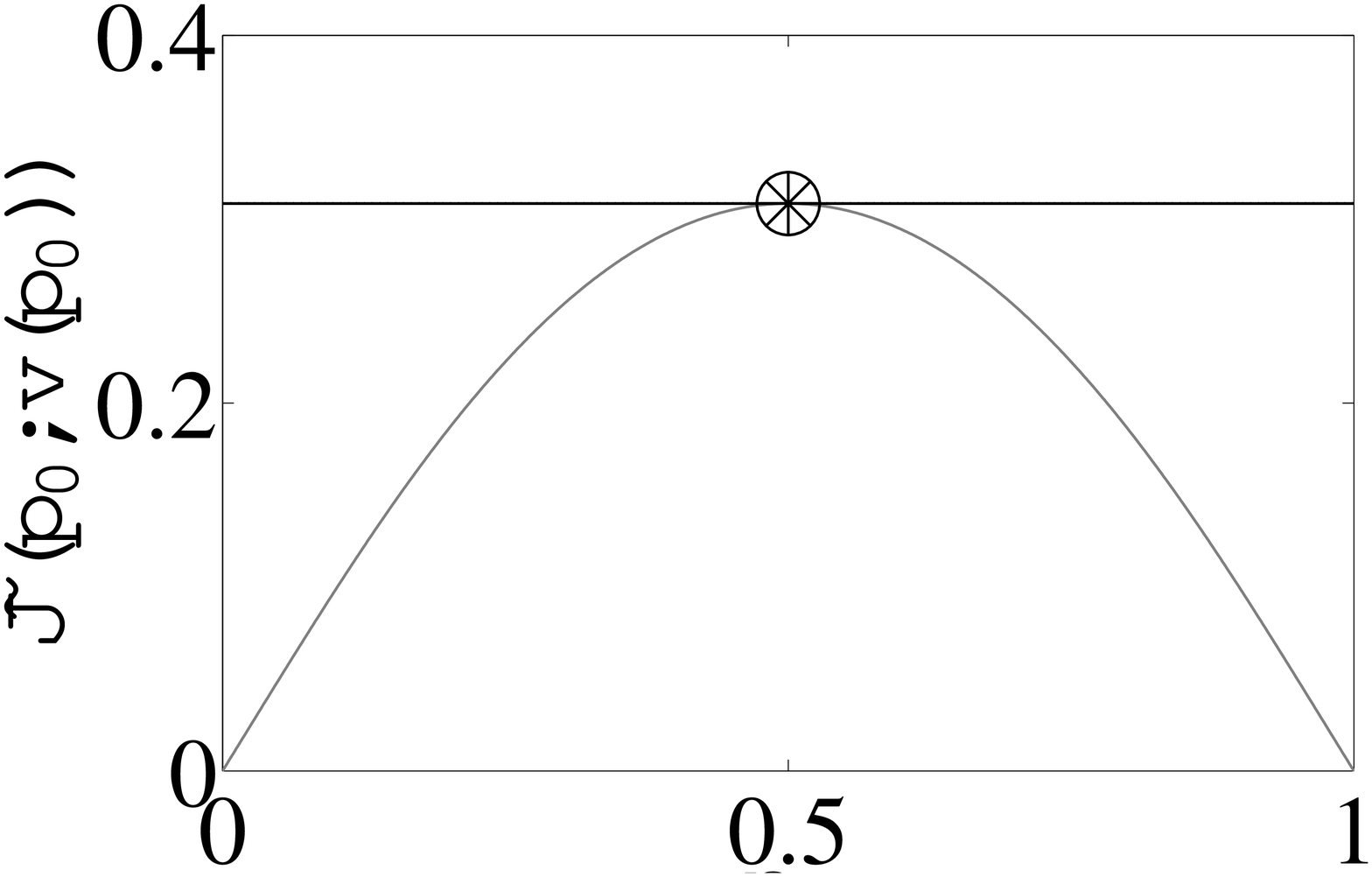} & \includegraphics[width=0.22\textwidth]{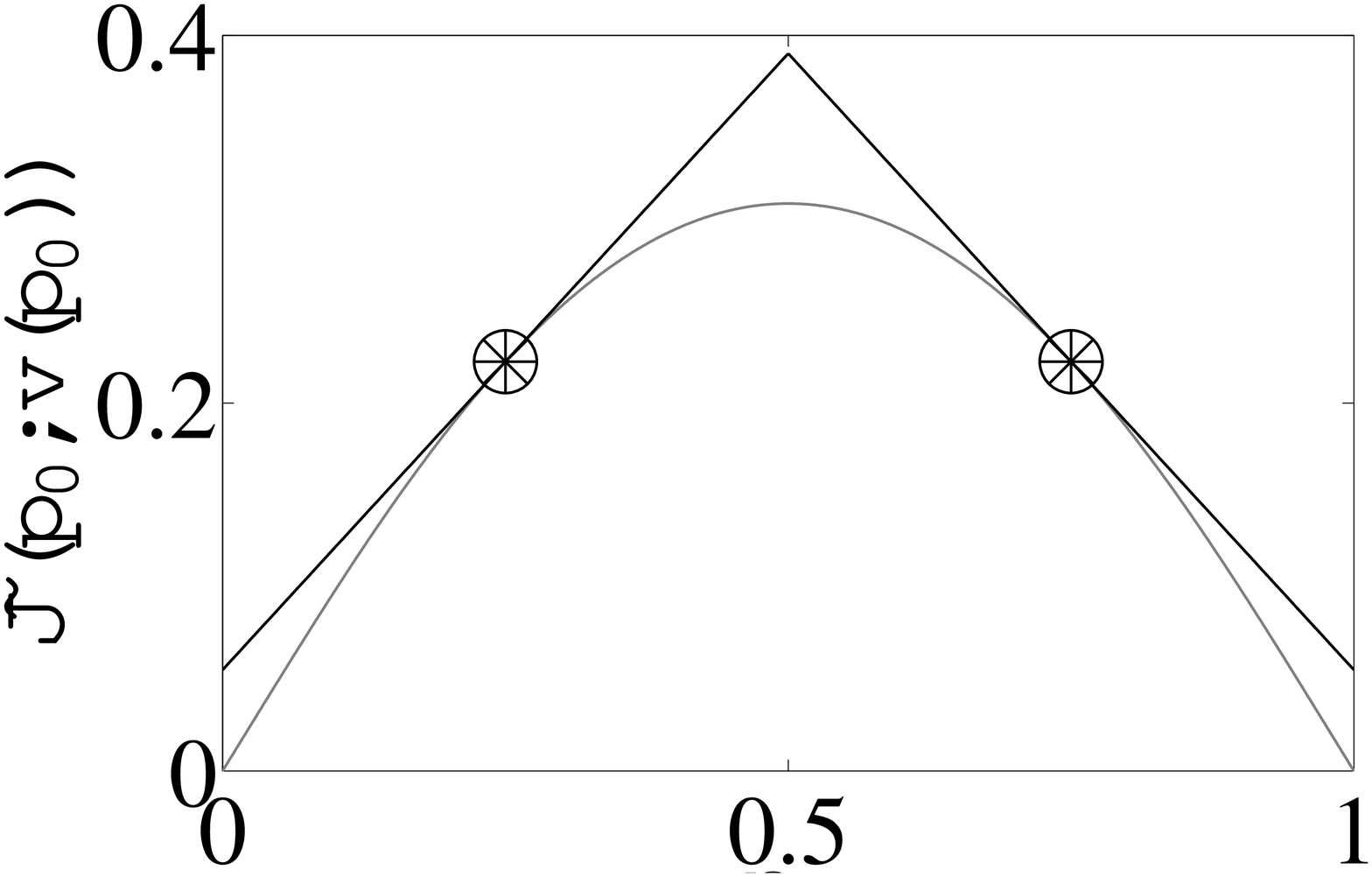} \\
			\footnotesize{(a)} & \footnotesize{(b)} \\
			\includegraphics[width=0.22\textwidth]{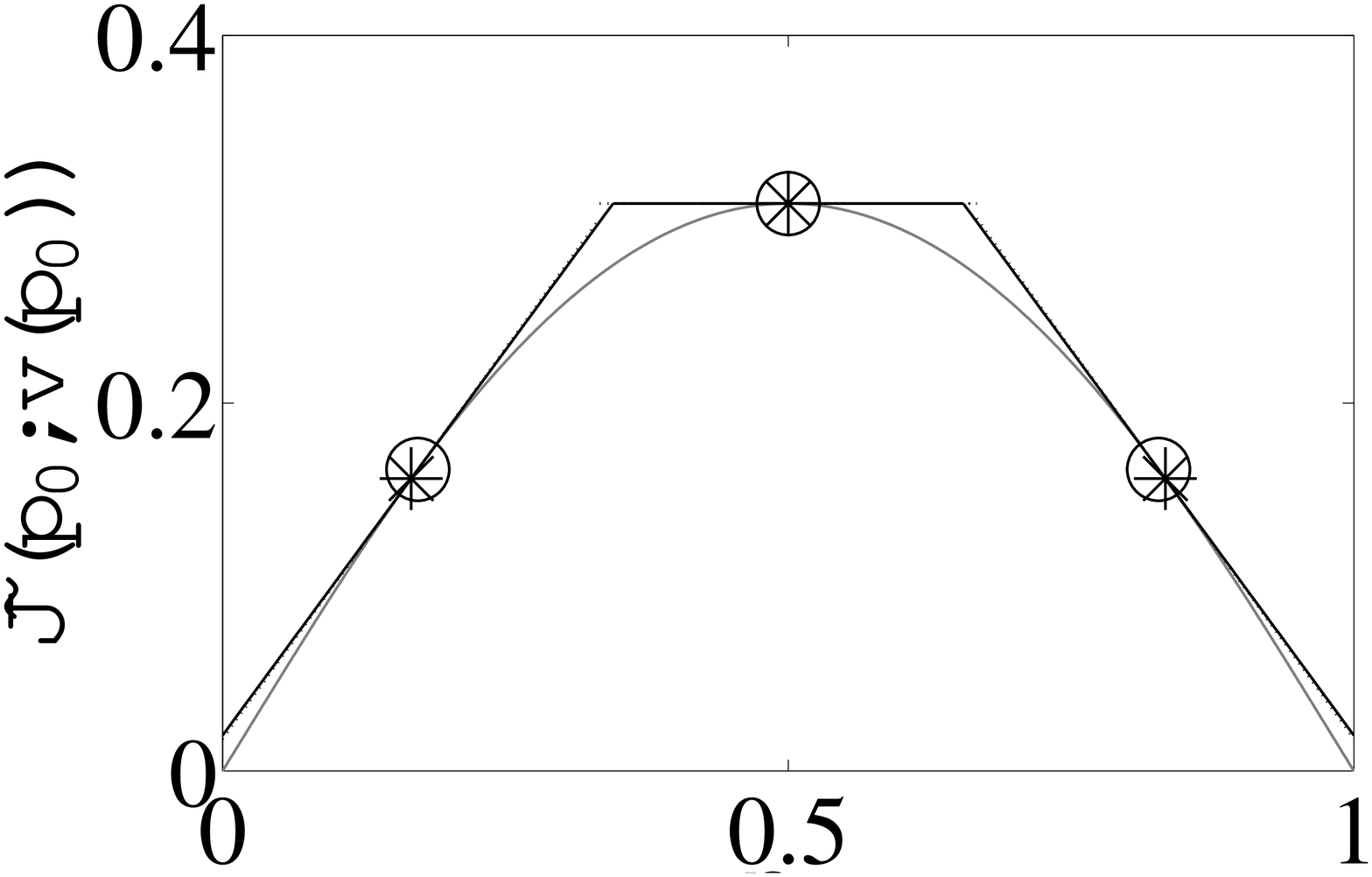} & \includegraphics[width=0.22\textwidth]{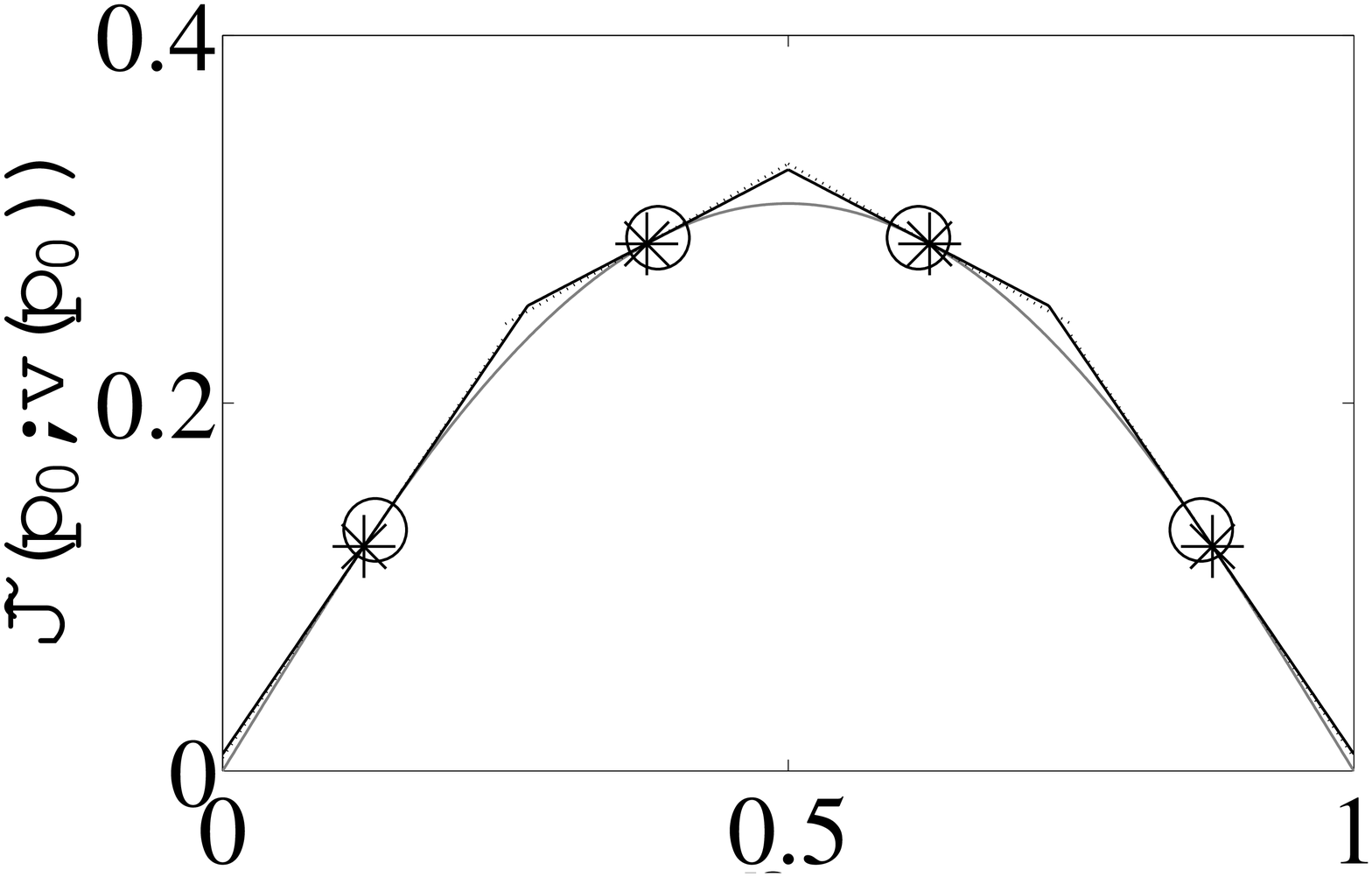} \\
			\footnotesize{(c)} & \footnotesize{(d)} \\
		\end{tabular}
	\end{center}
	\caption{Quantizers for uniformly distributed $P_0$ and Bayes costs $c_{10} = c_{01} = 1$.  $\tilde{J}(p_0,v_K(p_0))$ is plotted for (a) $K=1$, (b) $K=2$, (c) $K=3$, and (d) $K=4$; the markers, circle and asterisk for the MBRE-optimal and MAE-optimal quantizers respectively, are the representation points $\{a_k\}$.  The gray line is the unquantized Bayes risk $J(p_0)$.}
\label{fig:quantizers_uniform}
\end{figure}

\begin{figure}
	\begin{center}
		\includegraphics[width=0.48\textwidth]{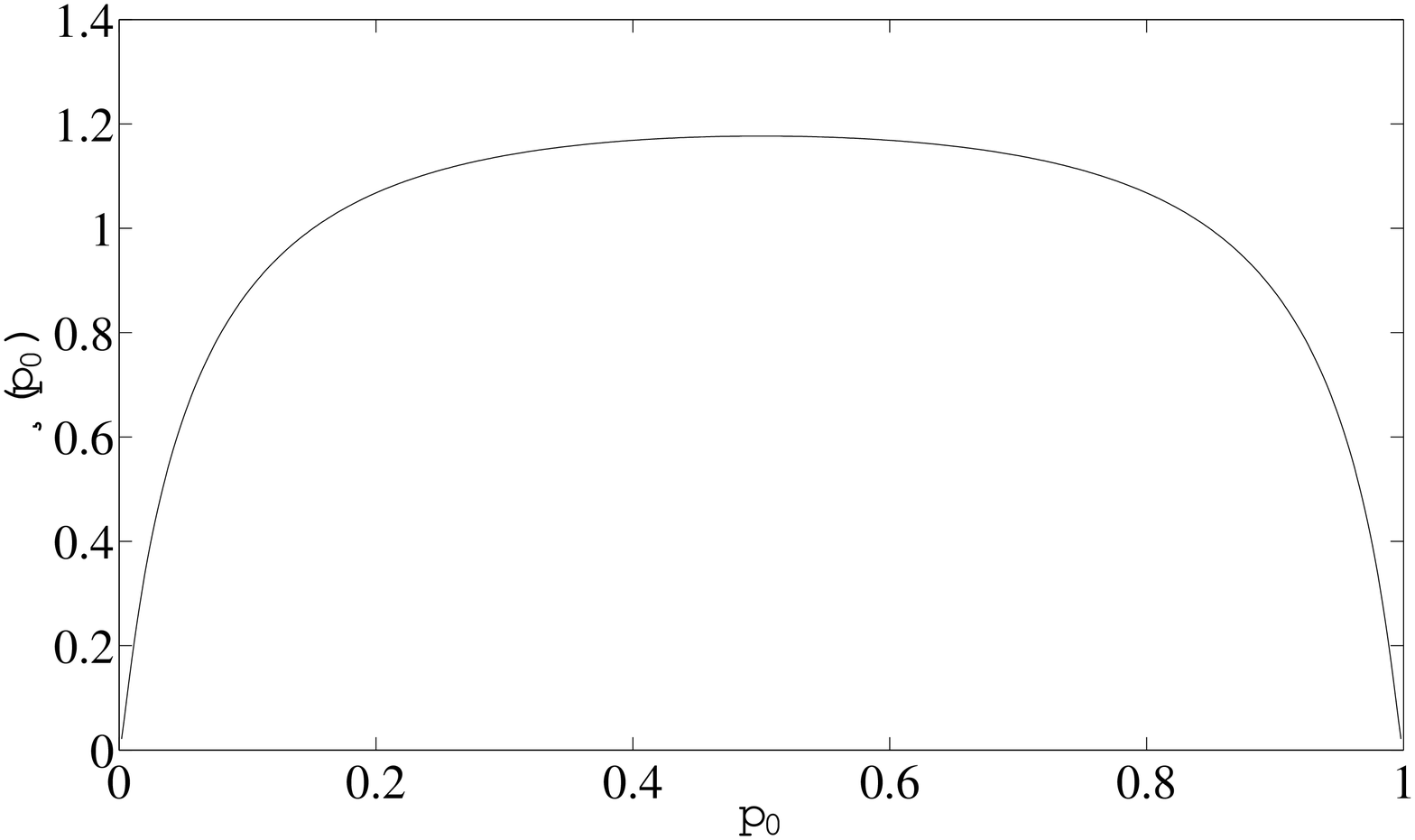}
	\end{center}
	\caption{Optimal MBRE point density for uniformly distributed $P_0$ and Bayes costs $c_{10} = c_{01} = 1$.}
\label{fig:pd_uniform}
\end{figure}

The performance of both quantizers is similar, but the MBRE-optimal quantizer always performs better or equally.  For $K=1,2$, the two quantizers are identical, as seen in Fig.~\ref{fig:quantizers_uniform}a-b.  
The plots in Fig.~\ref{fig:quantizers_uniform} show $\tilde{J}(p_0,v_K(p_0))$ as solid and dotted lines for the MBRE- and MAE-optimal quantizers respectively; the markers are the representation points.  The gray line is $J(p_0)$, the Bayes risk with unquantized prior probabilities.  For $K=3,4$, the representation points for the MBRE-optimal quantizer are closer to $p_0=\frac{1}{2}$ than the uniform quantizer.  This is because the area under the point density function $\lambda(p_0)$ shown in Fig.~\ref{fig:pd_uniform} is concentrated in the center.  
Each increment of $K$ is associated with a large reduction in Bayes risk.  There is a very large performance improvement from $K=1$ to $K=2$.  

\begin{figure}
	\begin{center}
		\includegraphics[width=0.48\textwidth]{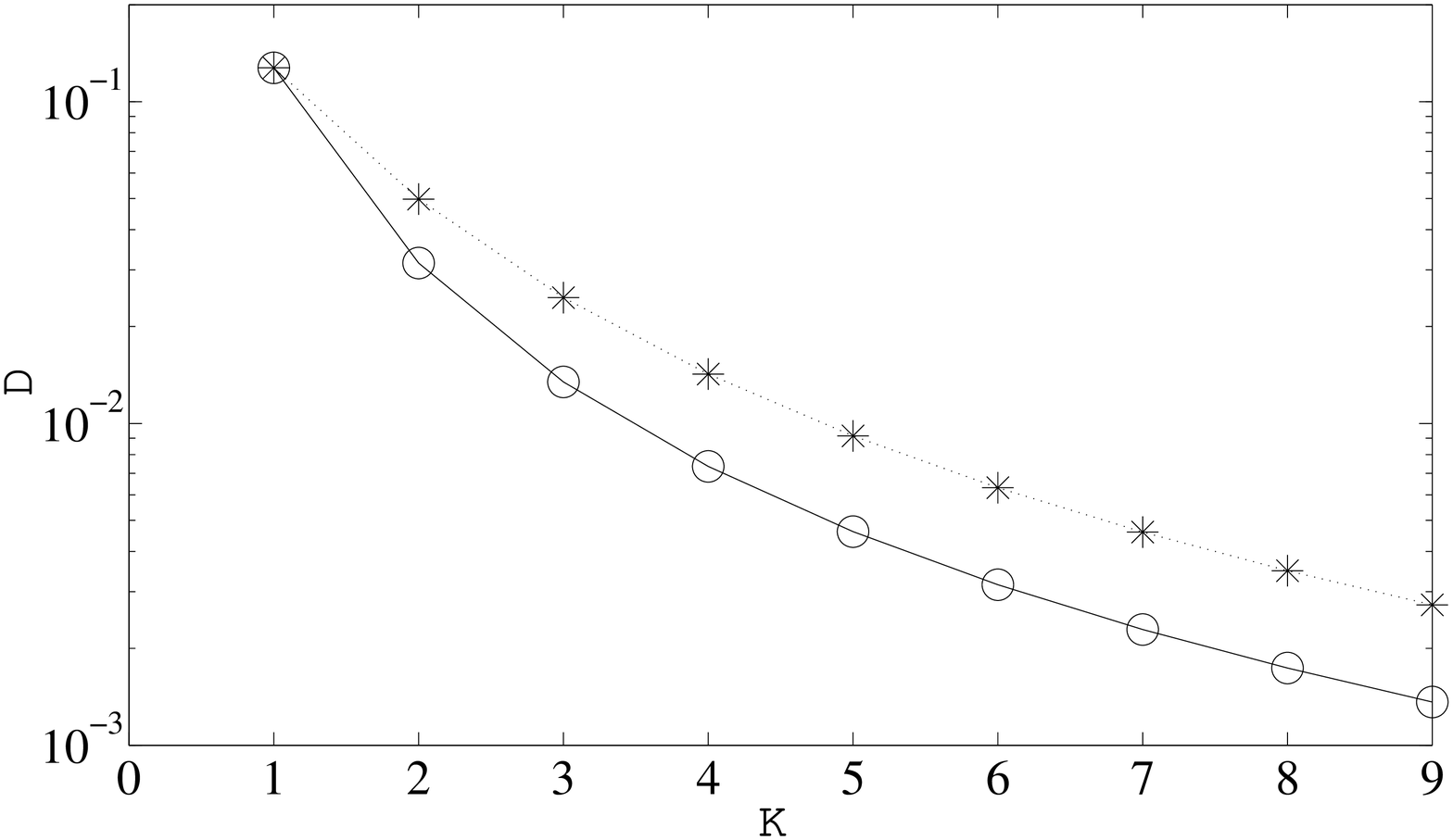}
	\end{center}
	\caption{MBRE for uniformly distributed $P_0$ and Bayes costs $c_{10} = 1, c_{01} = 4$ plotted on a logarithmic scale as a function of the number of quantization levels $K$; the solid line with circle markers is the MBRE-optimal quantizer and the dotted line with asterisk markers is the MAE-optimal uniform quantizer.}
\label{fig:MBRE_uniform_c01}
\end{figure}

\begin{figure}
	\begin{center}
		\includegraphics[width=0.48\textwidth]{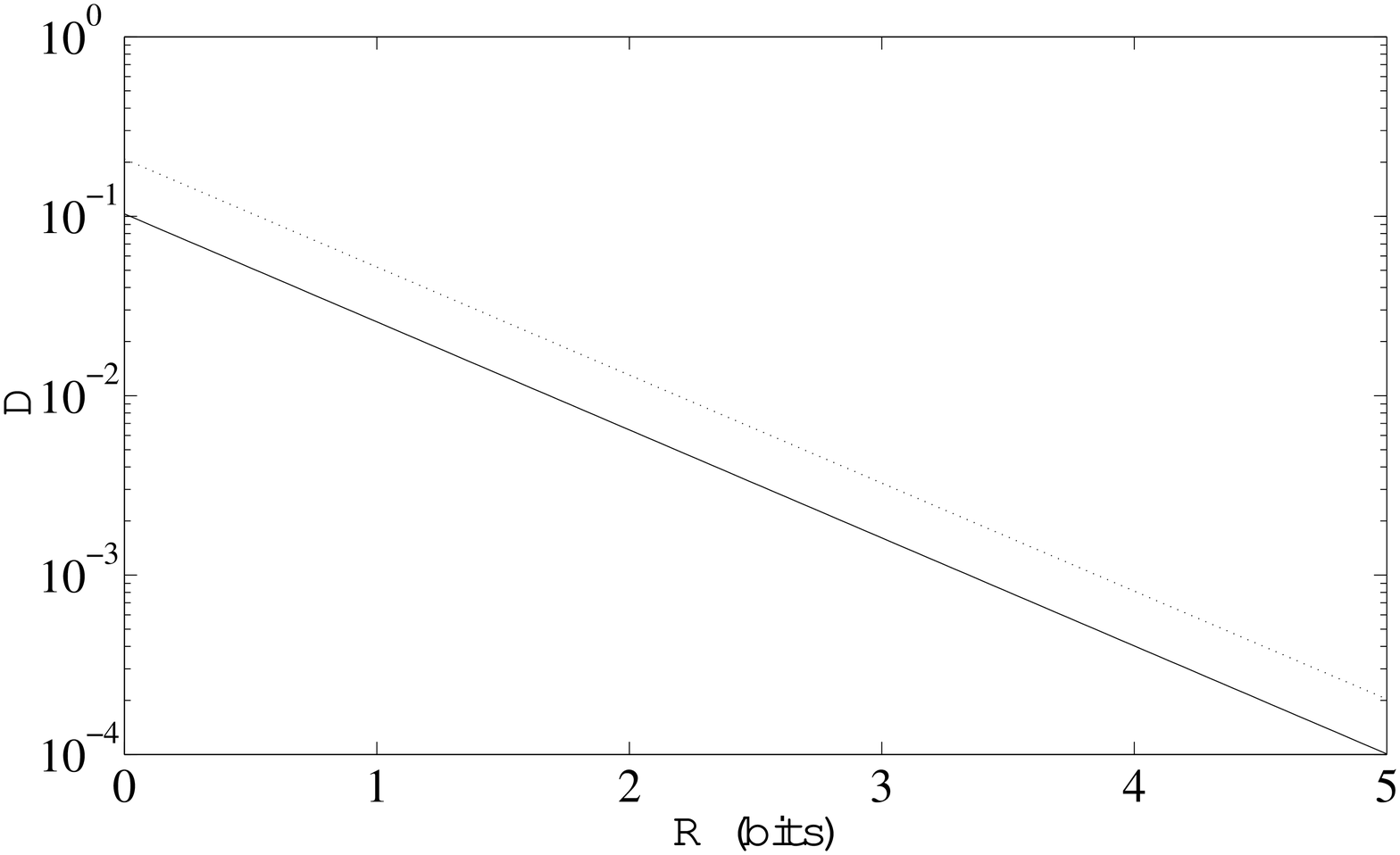}
	\end{center}
	\caption{High-rate approximation of distortion-rate function $D_L$ for uniformly distributed $P_0$ and Bayes costs $c_{10} = 1, c_{01} = 4$; the solid line is the MBRE-optimal quantizer and the dotted line is the MAE-optimal uniform quantizer.}
\label{fig:rd_uniform_c01}
\end{figure}

\begin{figure}
	\begin{center}
		\begin{tabular}{cc}
			\includegraphics[width=0.22\textwidth]{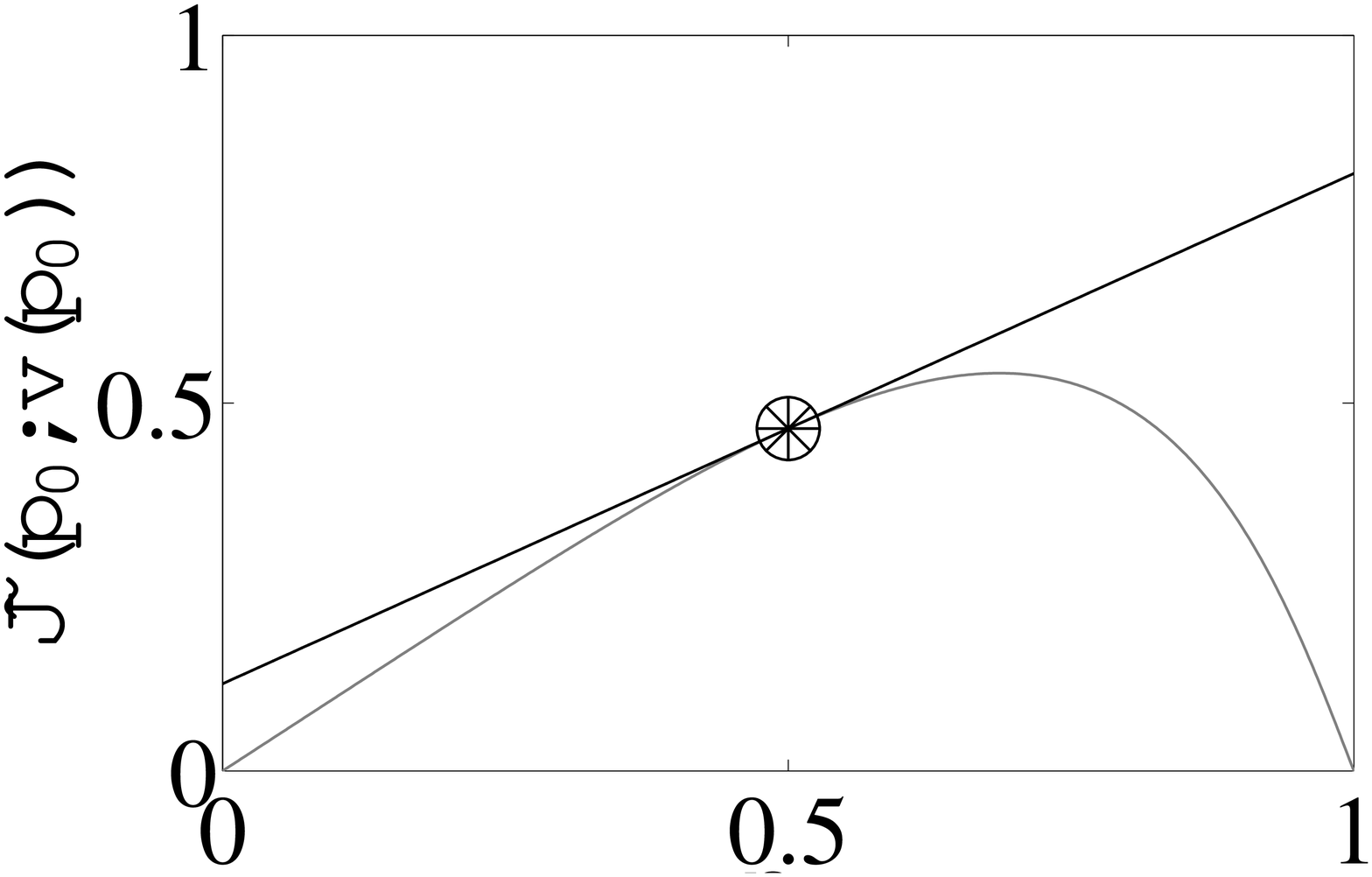} & \includegraphics[width=0.22\textwidth]{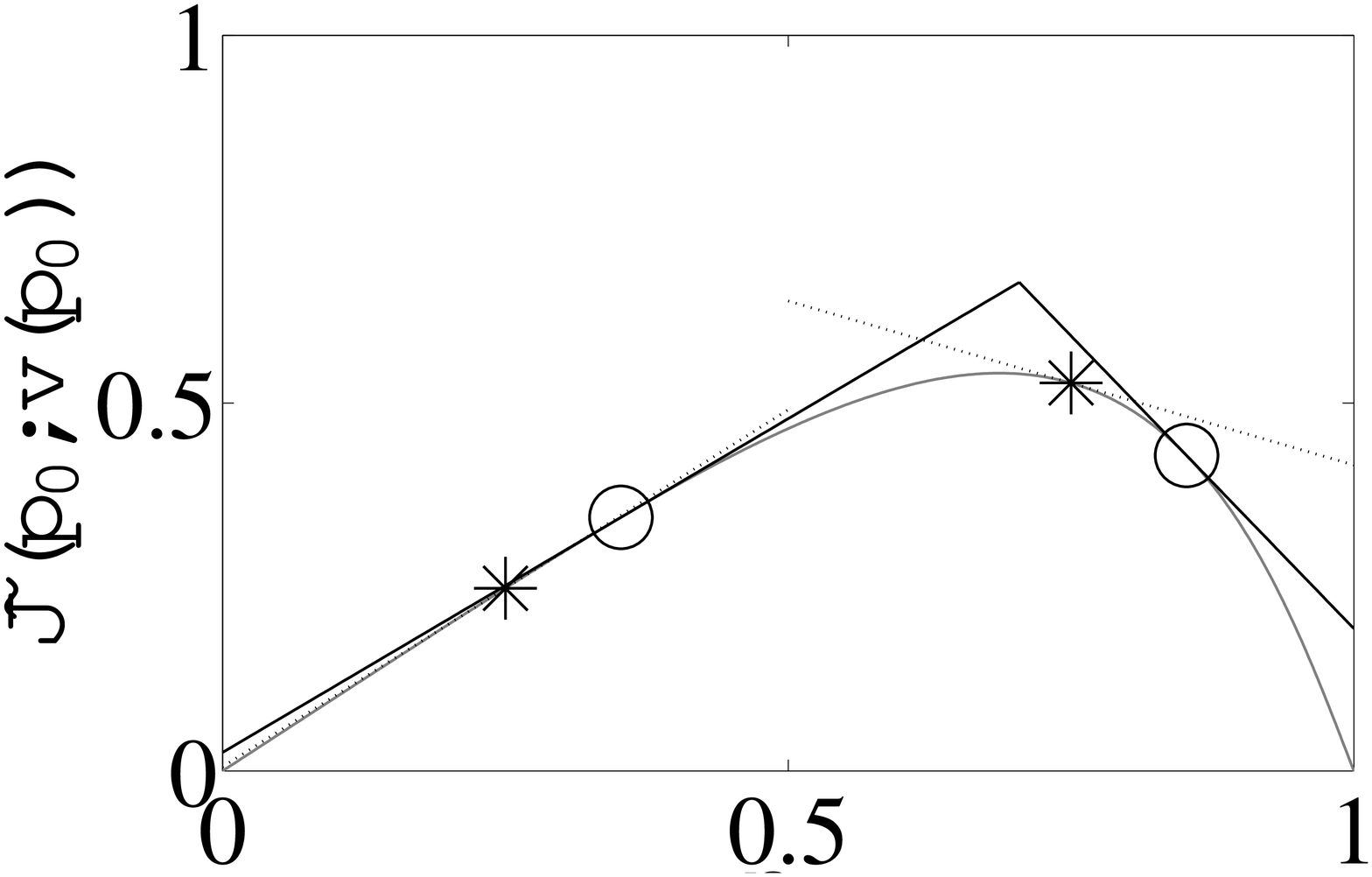} \\
			\footnotesize{(a)} & \footnotesize{(b)} \\
			\includegraphics[width=0.22\textwidth]{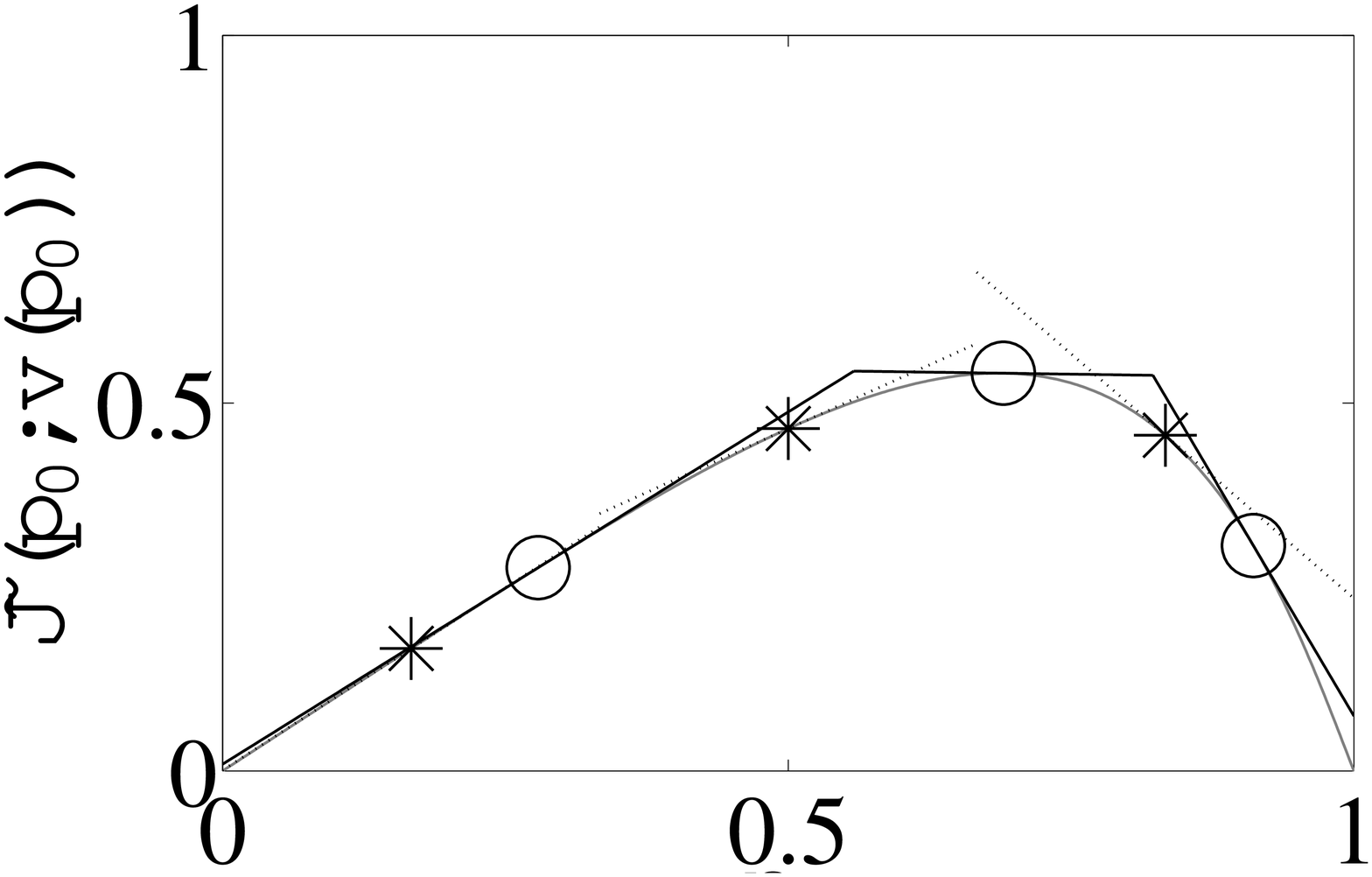} & \includegraphics[width=0.22\textwidth]{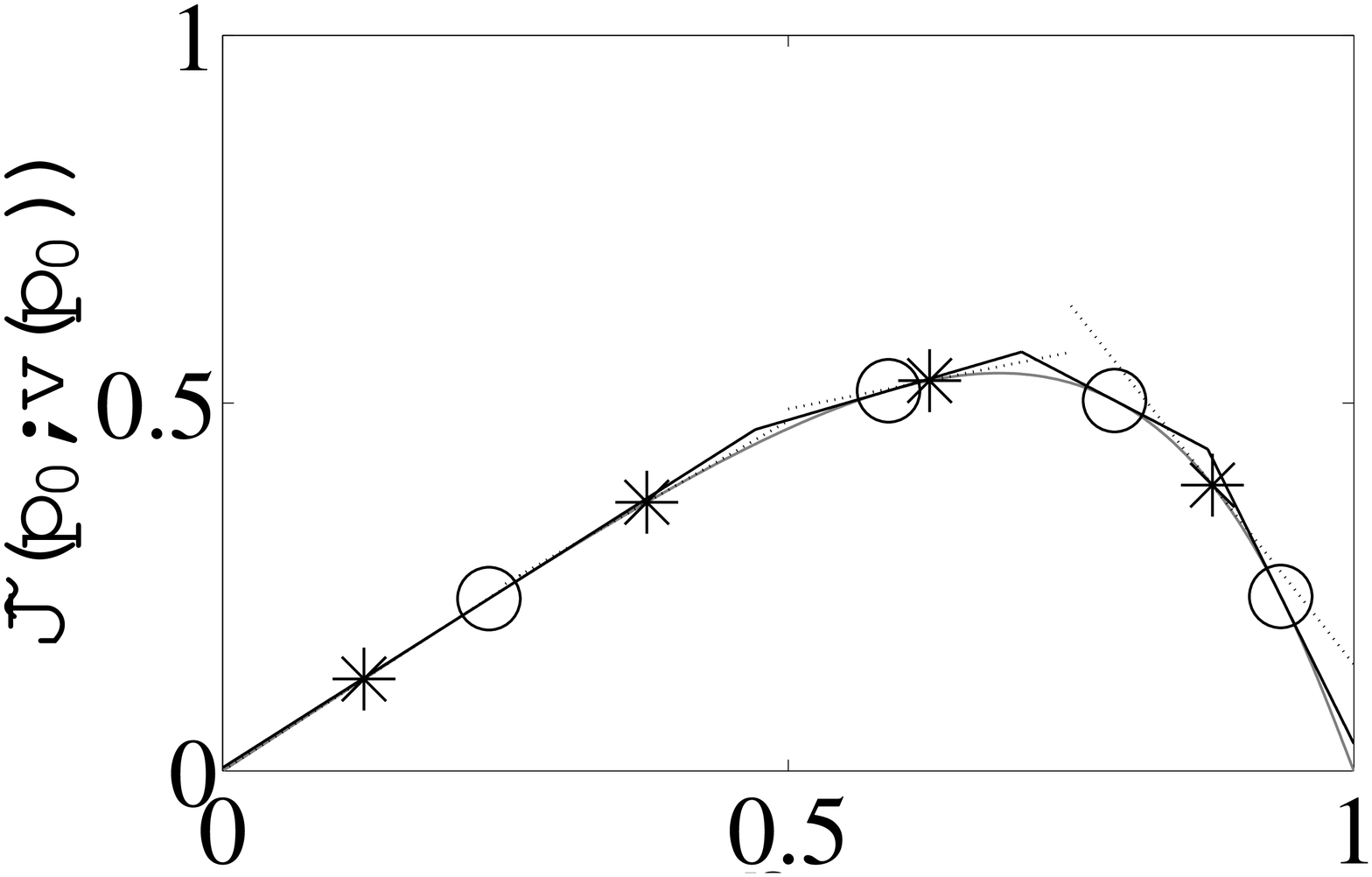} \\
			\footnotesize{(c)} & \footnotesize{(d)} \\
		\end{tabular}
	\end{center}
	\caption{Quantizers for uniformly distributed $P_0$ and Bayes costs $c_{10} = 1, c_{01} = 4$.  $\tilde{J}(p_0,v_K(p_0))$ is plotted for (a) $K=1$, (b) $K=2$, (c) $K=3$, and (d) $K=4$; the markers, circle and asterisk for the MBRE-optimal and MAE-optimal quantizers respectively, are the representation points $\{a_k\}$.  The gray line is the unquantized Bayes risk $J(p_0)$.}
\label{fig:quantizers_uniform_c01}
\end{figure}

\begin{figure}
	\begin{center}
		\includegraphics[width=0.48\textwidth]{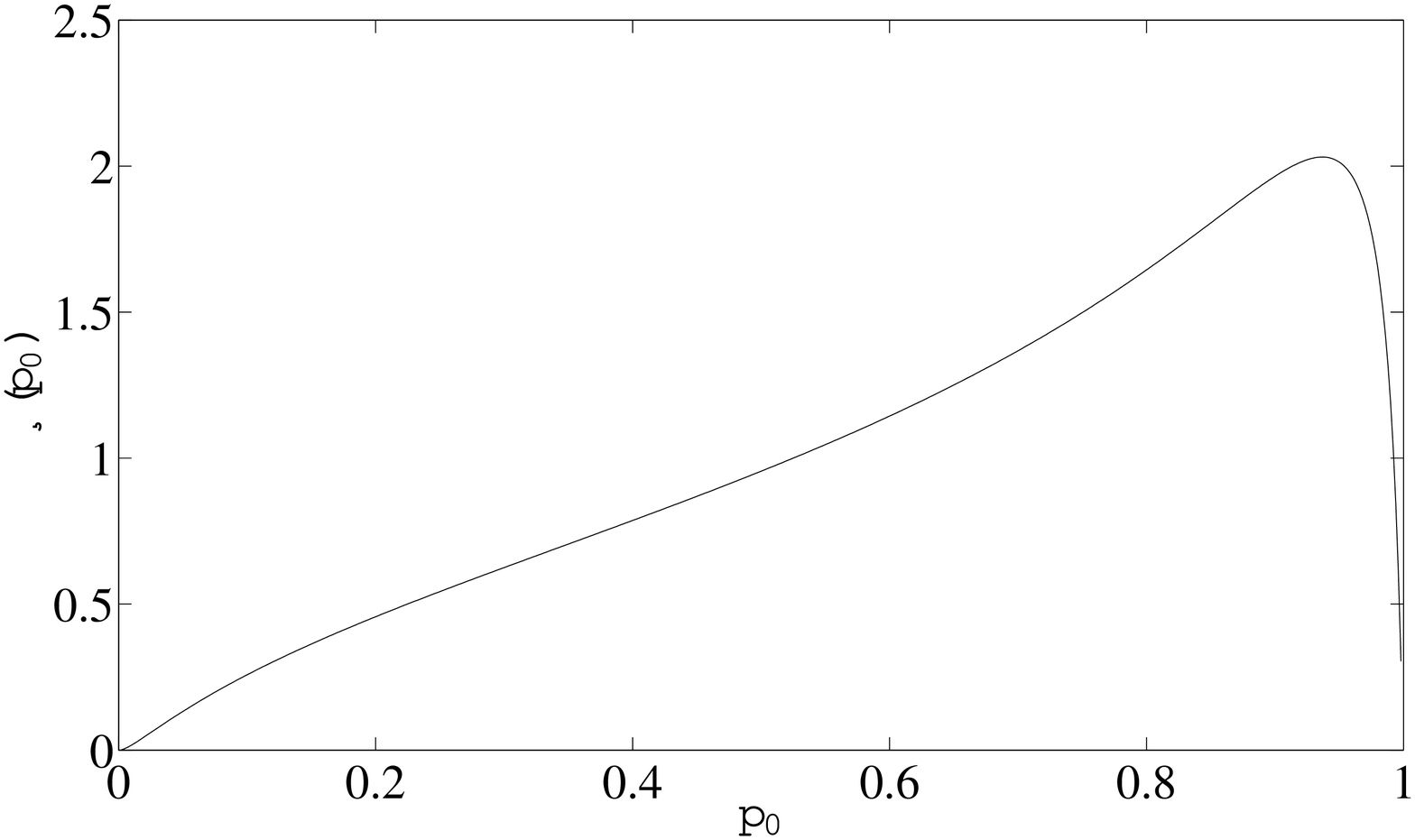}
	\end{center}
	\caption{Optimal MBRE point density for uniformly distributed $P_0$ and Bayes costs $c_{10} = 1, c_{01} = 4$.}
\label{fig:pd_uniform_c01}
\end{figure}

In Fig.~\ref{fig:MBRE_uniform_c01}, Fig.~\ref{fig:rd_uniform_c01}, Fig.~\ref{fig:quantizers_uniform_c01}, and Fig.~\ref{fig:pd_uniform_c01}, similar plots to those above are given for the case when the Bayes costs $c_{10}$ and $c_{01}$ are unequal.  
The unequal costs skew the Bayes risk function and consequently the representation point locations and point density function.  The difference in performance between the MBRE-optimal and MAE-optimal quantizers is greater in this example because the MAE-criterion cannot incorporate the Bayes costs, which factor into MBRE calculation.  

\subsection{Beta Distributed $P_0$}
\label{sec:example:beta}

\begin{figure}
	\begin{center}
		\includegraphics[width=0.48\textwidth]{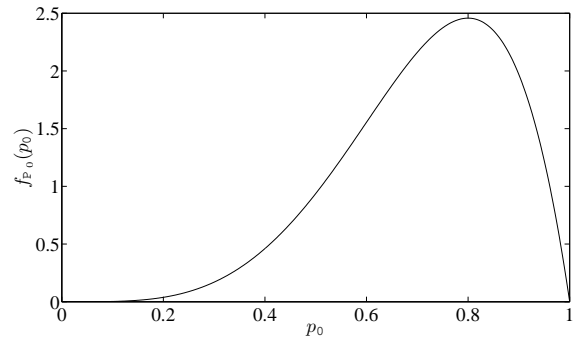}
	\end{center}
	\caption{The probability density function $f_{P_0}(p_0)$ for the Beta($5,2$) distribution.}
\label{fig:beta52_density}
\end{figure}

\begin{figure}
	\begin{center}
		\includegraphics[width=0.48\textwidth]{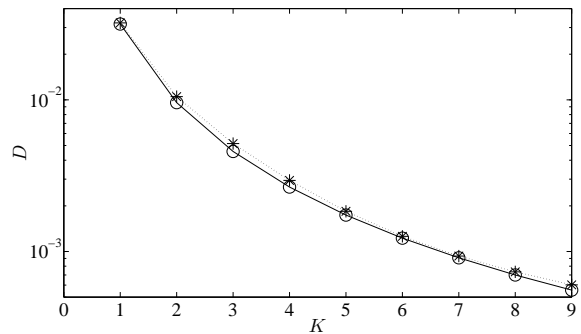}
	\end{center}
	\caption{MBRE for Beta($5,2$) distributed $P_0$ and Bayes costs $c_{10} = c_{01} = 1$ plotted on a logarithmic scale as a function of the number of quantization levels $K$; the solid line with circle markers is the MBRE-optimal quantizer and the dotted line with asterisk markers is the MAE-optimal uniform quantizer.}
\label{fig:MBRE_beta52}
\end{figure}

\begin{figure}
	\begin{center}
		\includegraphics[width=0.48\textwidth]{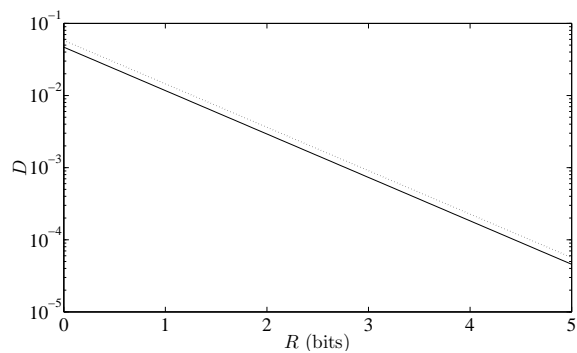}
	\end{center}
	\caption{High-rate approximation of distortion-rate function $D_L$ for Beta($5,2$) distributed $P_0$ and Bayes costs $c_{10} = c_{01} = 1$; the solid line is the MBRE-optimal quantizer and the dotted line is the MAE-optimal uniform quantizer.}
\label{fig:rd_beta52}
\end{figure}

Now, we look at a non-uniform distribution for $P_0$, in particular the Beta($5,2$) distribution.  The probability density function is shown in Fig.~\ref{fig:beta52_density}.  
The MBRE of the MBRE-optimal and MAE-optimal quantizers are in Fig.~\ref{fig:MBRE_beta52}.  
Here, there are also large improvements in performance with an increase in $K$.  The high-rate approximation to the distortion-rate function for this example is given in Fig.~\ref{fig:rd_beta52}.  

\begin{figure}
	\begin{center}
		\includegraphics[width=0.48\textwidth]{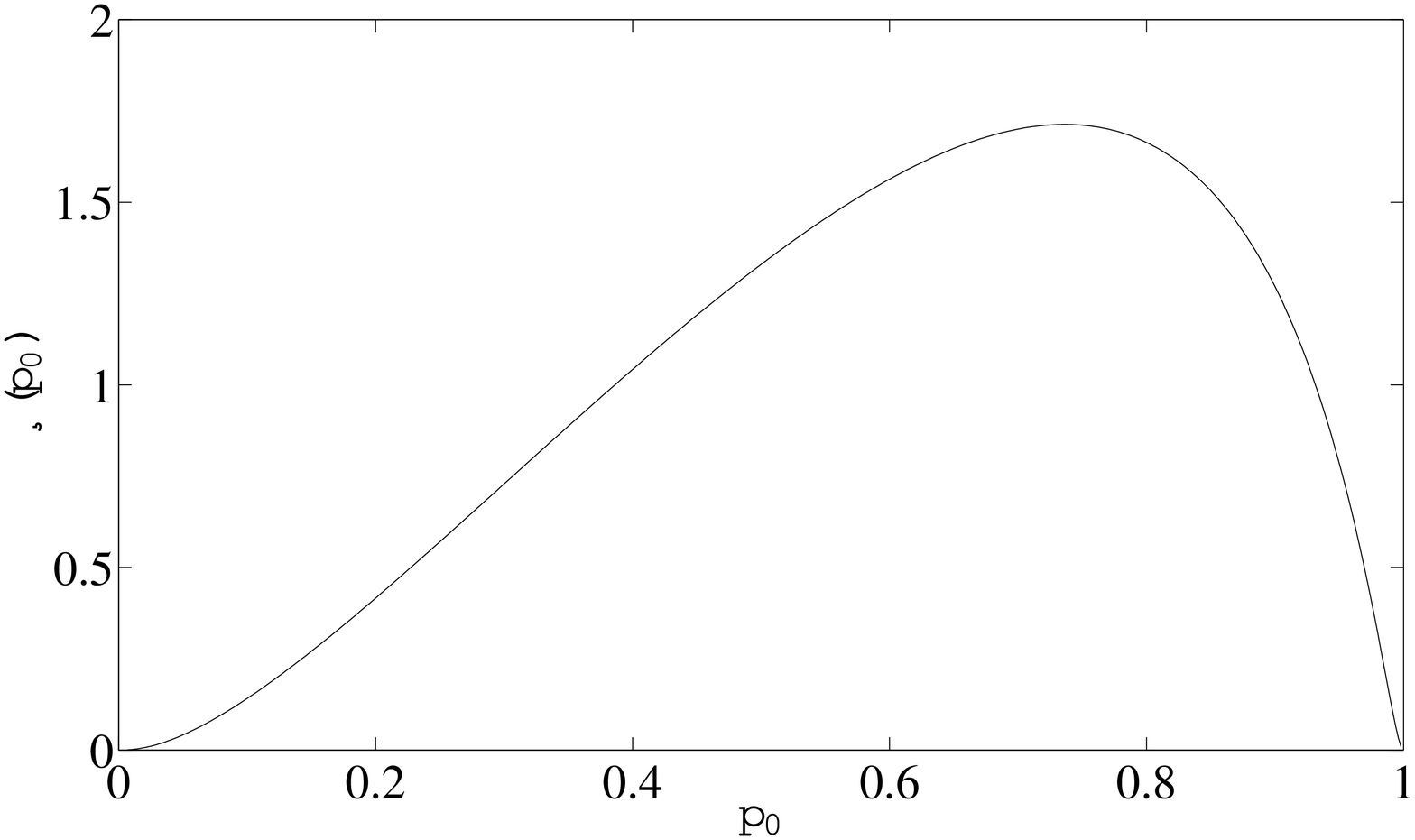}
	\end{center}
	\caption{Optimal MBRE point density for Beta($5,2$) distributed $P_0$ and Bayes costs $c_{10} = c_{01} = 1$.}
\label{fig:pd_beta52}
\end{figure}

\begin{figure}
	\begin{center}
		\begin{tabular}{cc}
			\includegraphics[width=0.22\textwidth]{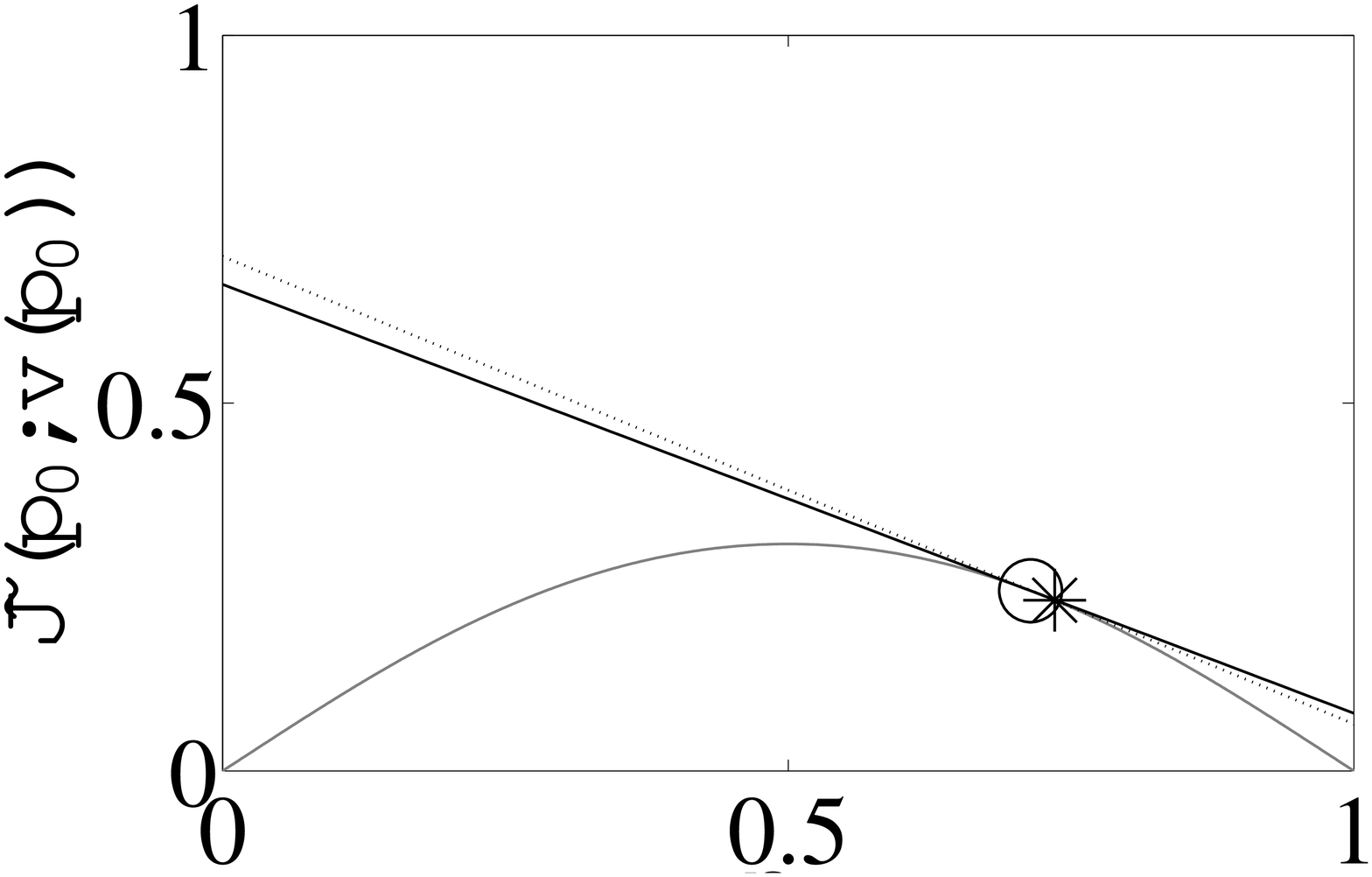} & \includegraphics[width=0.22\textwidth]{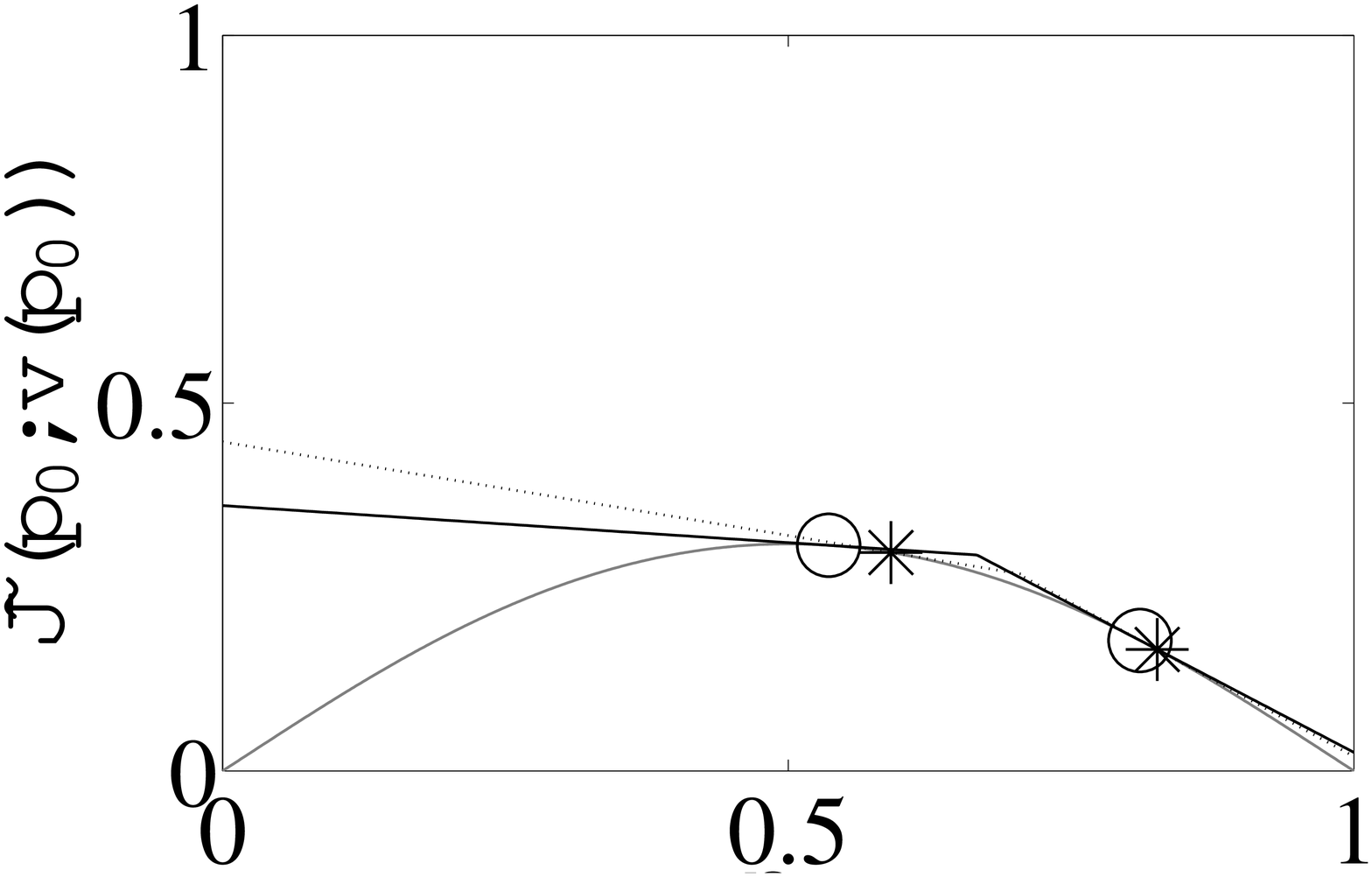} \\
			\footnotesize{(a)} & \footnotesize{(b)} \\
			\includegraphics[width=0.22\textwidth]{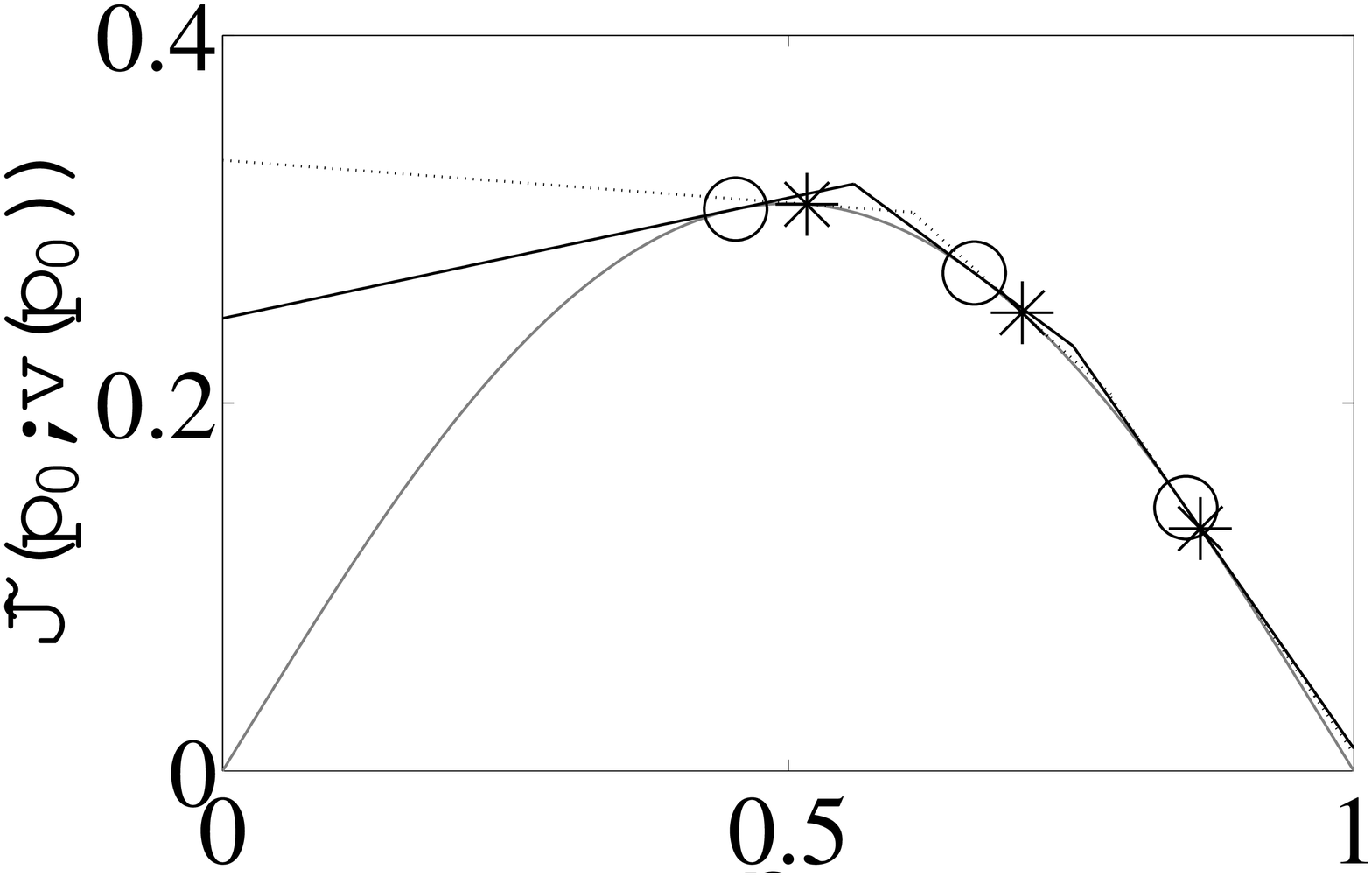} & \includegraphics[width=0.22\textwidth]{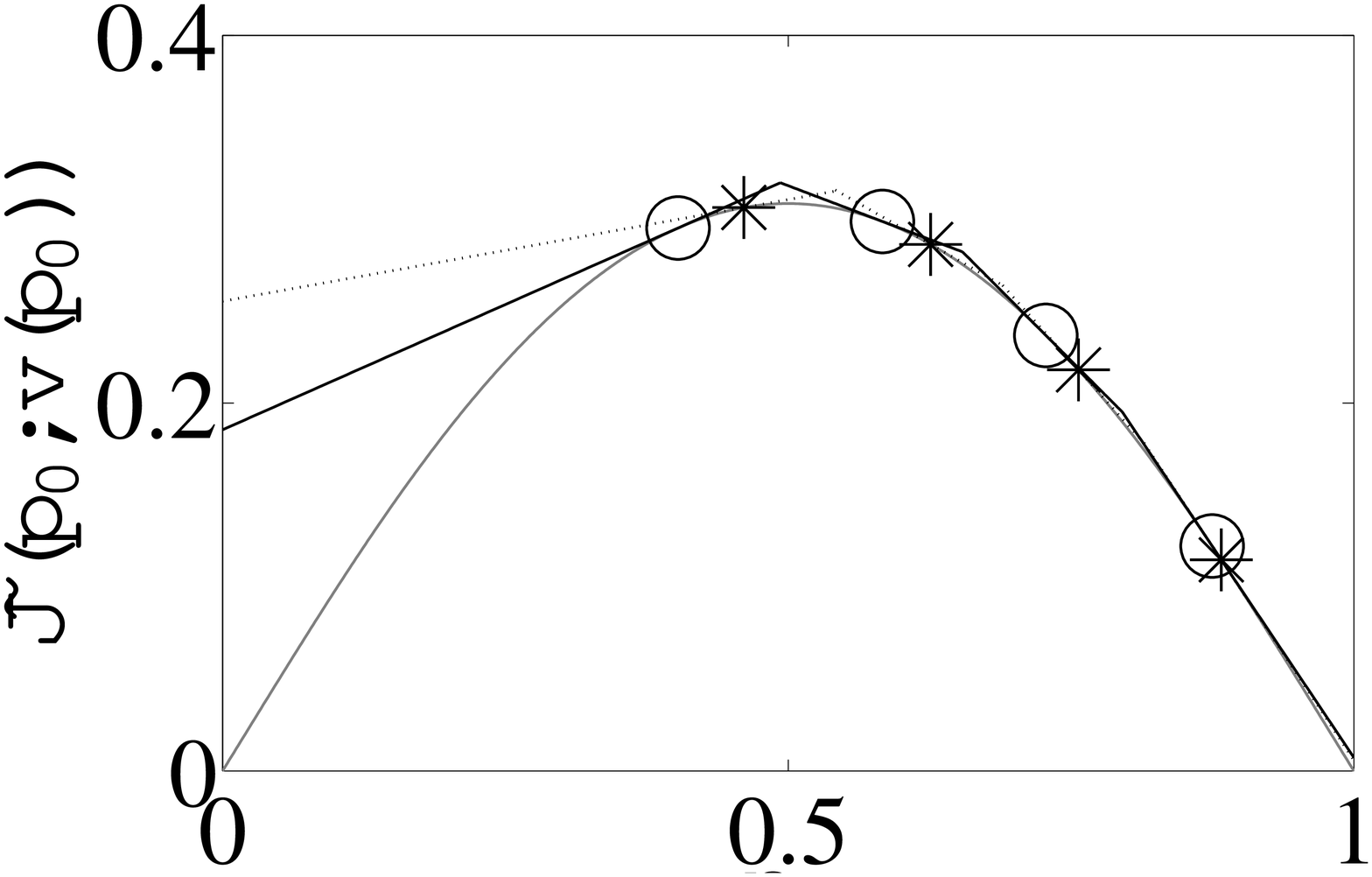} \\
			\footnotesize{(c)} & \footnotesize{(d)} \\
		\end{tabular}
	\end{center}
	\caption{Quantizers for Beta($5,2$) distributed $P_0$ and Bayes costs $c_{10} = 1, c_{01} = 4$.  $\tilde{J}(p_0,v_K(p_0))$ is plotted for (a) $K=1$, (b) $K=2$, (c) $K=3$, and (d) $K=4$; the markers, circle and asterisk for the MBRE-optimal and MAE-optimal quantizers respectively, are the representation points $\{a_k\}$.  The gray line is the unquantized Bayes risk $J(p_0)$.}
\label{fig:quantizers_beta52}
\end{figure}

The representation points $\{a_k\}$ are most densely distributed where $\lambda(p_0)$, plotted in Fig.~\ref{fig:pd_beta52}, has mass.  
In particular, more representation points are in the right half of the domain than in the left, as seen in Fig.~\ref{fig:quantizers_beta52}.    

\section{Implications on Human Decision Making}
\label{sec:impl}

In the previous sections, we formulated the minimum MBRE quantization problem and discussed how to find the optimal MBRE quantizer.  Having established the mathematical foundations of hypothesis testing with quantized priors, we may explore the implications of such resource-constrained decision making on human affairs.  

Let us consider the particular setting for human decision making mentioned in Section~\ref{sec:intro}: a referee determining whether a player has committed a foul or not using both his or her noisy observation and prior experience.  The fraction of plays in which a player commits a foul is that player's prior probability for $h_1$.  Over the population of players, there is a distribution of prior probabilities.  Also as mentioned in Section~\ref{sec:intro}, human decision makers categorize into a small number of categories due to limitations in information processing capacity \cite{Miller1956}.  Decisions by humans may be modeled via quantization of the distribution of prior probabilities and the use of the quantization level centroid of the category in which a player falls as the prior probability when performing hypothesis testing on that player's action.  

Therefore, a referee will do a better job with more categories rather than fewer.  A police officer confronting an individual with whom he or she has prior experience will make a better decision if he or she has the mental categories `probably violent,' `possibly violent or nonviolent,' and `probably nonviolent,' versus just `violent' and `nonviolent.'  Similarly, a doctor will have a smaller probability of error when interpreting a blood test if he or she knows the prior probability of the test turning out positive for many categorizations of patients rather than just one for the entire population at large.  Additional examples could be given for a variety of decision-making tasks.  Implications of this sort are not surprising.  However, when one additional component is added to the decision-making scenario, some fairly interesting implications arise.  Next, we look at the case when the quantization of two distinct populations is done separately.  

We discuss mathematically unavoidable consequences of quantized prior hypothesis testing when quantizing the prior probability for a minority population and the prior probability for a majority population separately, while taking identical prior probability distributions of the two populations $f_{P_0}(p_0)$.  Although majority and minority populations can be defined along any socially observable dimension, such as gender or age \cite{AkerlofK2000}, for ease of exposition we use race, and more specifically use `white' and `black' to denote the two populations.  Although there is some debate in the social cognition literature \cite{MacraeB2000}, it is thought that race and gender categorization is essentially automatic, particularly when a human actor lacks the motivation, time, or cognitive capacity to think deeply.  

We can extend the definition of MBRE to two populations as:
\begin{equation}
\label{eq:2mbre}
	D^{(2)} = \tfrac{w}{w+b}E[\tilde{J}(P_0,v_{K_w}(P_0))] + \tfrac{b}{w+b}E[\tilde{J}(P_0,v_{K_b}(P_0))] - E[J(P_0)],
\end{equation}  
where $w$ is the number of whites encountered, $b$ is the number of blacks encountered,\footnote{One might assume that $w$ and $b$ are simply the number of whites and blacks in the general population, however these numbers should actually be based on the social interaction pattern of the decision maker.  Due to segregation in social interaction, see e.g.~\cite{EcheniqueF2007} and references therein, there is greater intra-population interaction than inter-population interaction.  The decision maker has more training data from intra-population interaction.} $K_w$ is the number of points in the quantizer for whites, and $K_b$ is the number of points in the quantizer for blacks.  In order to find the optimal allocation of the total quota of representation points $K_t = K_w + K_b$, we minimize $D^{(2)}$ for all $K_t-1$ possible allocations and choose the best one; more sophisticated algorithms developed for bit allocation to subbands in transform coding may also be used \cite{ShohamG1988}.  

Fryer and Jackson have previously suggested that it is better to allocate more representation points to the majority population than to the minority population \cite{FryerJ2008}.  With two separate scalar quantizers, but a single size constraint, optimizing $D^{(2)}$ over $v_{K_w}(\cdot)$ and $v_{K_b}(\cdot)$ yields the same result.  Due to the monotonicity result in Sec.~\ref{sec:lowrate:convergence}, the MBRE for members of the minority group is greater than that for the majority group.  

Assuming white decision makers have $w > b$ and black decision makers have $b > w$, analysis of quantized prior Bayesian hypothesis testing predicts that there should be own-race bias in decision making.  This prediction is in fact born out experimentally.  A large body of literature in face recognition shows exactly the predicted own race bias effect, observed colloquially as ``they [other-race persons] all look alike.''  In particular, both parts of the Bayes risk, $p_E^{\text{I}}$ and $p_E^{\text{II}}$ increase when trying to recognize members of the opposite population \cite{MeissnerB2001}.  Verification of own race bias in face recognition is due to laboratory experimentation, however similar effects have also been observed in natural experiments through econometric studies.

It has been found that the addition of police officers of a given race is associated with an increase in the number of arrests of suspects of a different race but has little impact on same-race arrests.  The effect is more pronounced for minor offenses where the prior probability presumably plays a bigger role than the measurement \cite{DonahueL2001}.  There are similar own-race bias effects in the decision by police to search a vehicle during a traffic stop \cite{AntonovicsK2004}, in the decision of human resource professionals to not hire \cite{StollRH2004}, and in the decision of National Basketball Association (NBA) referees to call a foul \cite{PriceW2007}.  The rate of searching, the rate of not hiring, and the rate of foul calling are all greater when the decision-maker is of a different race than the driver, applicant, and player, respectively.  A major difficulty in interpreting these econometric studies, however, is that the ground truth is not known.  Higher rates may be explained by either greater $p_E^{\text{I}}$ or smaller $p_E^{\text{II}}$.  

Since ground truth is lacking in econometric studies, it is not clear how to interpret a finding that white referees call more fouls on black players and that black referees call more fouls on white players.  This phenomenon cannot simply be explained by a larger probability of decision error.  The Bayes risk must be teased apart into its constituent parts and the Bayes costs must be examined in detail.  

The measurable quantity in an econometrics study is the probability that a foul is called:
\begin{equation}
\label{eq:foulrate}
	\Pr[\hat{H}_K = h_1] = 1 - p_0 + p_0p_E^{\text{I}}(v_K(p_0)) - (1-p_0)p_E^{\text{II}}(v_K(p_0)).
\end{equation}
Looking at the average performance of a white referee over the populations of black and white players, we compare the expected foul rates on whites and blacks ($K_b<K_w$):
\begin{equation}
\label{eq:foulratediff}
	\Delta = E\left[\Pr[\hat{H}_{K_b} = h_1] - \Pr[\hat{H}_{K_w} = h_1]\right].
\end{equation}
If this discrimination quantity $\Delta$ is greater than zero, then the white referee is calling more fouls on blacks.  If $\Delta$ is less than zero, then the referee is calling more fouls on whites.  The $\Delta$ expression may be written as:
\begin{multline}
\label{eq:foulratediff_s}
	\Delta(c_{10},c_{01}) = E[p_0p_E^{\text{I}}(v_{K_b}(p_0)) - (1-p_0)p_E^{\text{II}}(v_{K_b}(p_0))]\\ - E[p_0p_E^{\text{I}}(v_{K_w}(p_0)) - (1-p_0)p_E^{\text{II}}(v_{K_w}(p_0))].
\end{multline}
The dependence of $\Delta$ on $c_{10}$ and $c_{01}$ is explicit on the left side of \eqref{eq:foulratediff_s} and is implicit in the error probabilities on the right side.  The value of $\Delta$ also depends on the unquantized prior distribution $f_{P_0}(p_0)$, the measurement model, and the quantizer.  

If the prior distribution and measurement model are fixed, and the MBRE-optimal quantizer used, we find that the regions in the $c_{10}$-$c_{01}$ plane where a white referee would call more fouls on blacks and where a white referee would call more fouls on whites are half-planes.  For the uniform prior $f_{P_0}(p_0)$, the dividing line between the two regions is exactly $c_{01} = c_{10}$.  For the Beta($5$,$2$) prior, the dividing line is $c_{01} = mc_{10}$, where $m>1$.  

Using the division of the $c_{10}$-$c_{01}$ plane into two parts, we can now interpret the econometric findings in the NBA referee study \cite{PriceW2007} and related results \cite{DonahueL2001,AntonovicsK2004,StollRH2004}.  The NBA race bias observations can be generated from the quantized prior hypothesis testing model only if the Bayes risk error has costs $c_{01} > c_{10}$ for a uniform prior or costs $c_{01} \gg c_{10}$ for a Beta($5$,$2$) prior.  The choice of Bayes costs with $c_{01}$ greater than $c_{10}$ implies that a referee can tolerate more instances of calling fouls on plays that are not fouls rather than the opposite.  This assignment of costs has been called the precautionary principle in some contexts.  Very simply, the precautionary principle states ``better safe than sorry.''  

Taken together, the hypothesis testing with quantized priors model, the phenomenon of racial segregation \cite{EcheniqueF2007}, and results from econometric studies \cite{PriceW2007,DonahueL2001,AntonovicsK2004,StollRH2004} suggest that referees, police officers, and human resources professionals all follow the precautionary principle.

\section{Conclusion and Future Work}
\label{sec:conc}

We have looked at Bayesian hypothesis testing when there is a distribution of prior probabilities, but the decision maker may only use a quantized version of the true prior probability in designing a decision rule.  Considering the problem of finding the optimal quantizer for this purpose, we have defined a new fidelity criterion based on the Bayes risk function.  For this criterion, MBRE, we have determined the conditions that an optimal quantizer satisfies and worked through a high-rate approximation to the distortion.  $M$-ary hypothesis testing with $M>2$ requires vector quantization rather than scalar quantization, but determining the Lloyd-Max conditions and high-rate theory is no different conceptually due to the geometry of the Bayes risk function and mismatched Bayes risk function.  For the $M$-ary hypothesis testing case, a multivariate distribution such as the $M$-dimensional Dirichlet distribution \cite{Fine2006} is needed for $f_{\textsc{\mbox{\boldmath$P$}}}(\mbox{\boldmath$p$})$.  Previous, though significantly different, work on quantization for hypothesis testing was unable to directly minimize the Bayes risk, as was accomplished in this work.  

The mathematical theory of quantized prior hypothesis testing formulated here leads to a generative model of discriminative behavior  when combined with theories of social cognition and empirical facts about social segregation.  This biased decision making arises despite having identical distributions for different populations and despite no malicious intent on the part of the decision maker.  We also discussed how the choice of Bayes costs affects detection probabilities; in particular, the precautionary principle leads to a higher detection probability for the opposite race, whereas a more optimistic view leads to a higher detection probability for the own race.  Such a phenomenon of pessimistic or optimistic attitude fundamentally altering the nature of discrimination seems not to have been described before.  
Discrimination on the basis of race, gender, and other socially observable characteristics has been a troublesome social problem, but appears to be a permanent artifact of the automaticity of classification and the finite human capacity for information processing.  

There are many avenues along which to extend this work, such as dealing with decentralized detection and classification (with possible implications on jury decisions and elections), which may become game theoretic; consideration of additional noise before or after quantization of the prior probabilities; or the development of successively refinable quantizers (for decision makers that possess a memory hierarchy).  One can also consider a restricted class of quantizers rather than considering optimal quantization.  Such restriction may model further cognitive constraints on human decision makers.  In particular, Fryer and Jackson have suggested a heuristic algorithm for quantizer design based on splitting groups \cite{FryerJ2008}, which is a rediscovery of the tree-structured vector quantizer (TSVQ) design algorithm given in \cite[Fig.~20]{MakhoulRG1985}.  Beyond \cite{MakhoulRG1985}, there has been much recent development in the theory of TSVQ performance and recursive partitioning, which may prove useful.  

For the quantizer with $K=1$, an alternative to the MBRE-optimal representation point:
\begin{displaymath}
	\mbox{\boldmath$a$}_{\text{MBRE}}^* = \arg\min_{\mbox{\boldmath$a$}}\left\{\int \tilde{J}(\mbox{\boldmath$p$},\mbox{\boldmath$a$}) f_{\textsc{\mbox{\boldmath$P$}}}(\mbox{\boldmath$p$}) d\mbox{\boldmath$p$}\right\}
\end{displaymath}
is the min-max hypothesis testing representation point:
\begin{displaymath}
	\mbox{\boldmath$a$}_{\text{min-max}}^* = \arg\min_{\mbox{\boldmath$a$}}\left\{\max_{\mbox{\boldmath$p$}} \tilde{J}(\mbox{\boldmath$p$},\mbox{\boldmath$a$}) \right\},
\end{displaymath}
which is only equivalent in special cases.  A distribution on the prior probabilities is needed to specify $\mbox{\boldmath$a$}_{\text{MBRE}}^*$, but not to specify $\mbox{\boldmath$a$}_{\text{min-max}}^*$.  One may consider extending the min-max idea to $K>1$.  This would involve an approach related to $\epsilon$-entropy \cite[Sec.~6.1.2]{Berger1971} and finding a cover for the unit simplex by $K$ sets of the form $\mathcal{R}_k = \{\mbox{\boldmath$p$}|\tilde{J}(\mbox{\boldmath$p$},\mbox{\boldmath$a$}_k) \le D\}$, where all $\mbox{\boldmath$p$}$ in $\mathcal{R}_k$ map to $\mbox{\boldmath$a$}_k$ and $D$ is the same for all $\mathcal{R}_k$.  

The general theme of machine learning for the explicit purpose of hypothesis testing, within which this work falls, is receiving increasing attention; framing the hypothesis testing scenario discussed here in terms of probabilistic graphical models of categorization, e.g.~the latent Dirichlet allocation model \cite{BleiNJ2003} and the hierarchical Dirichlet process mixture model \cite{TehJBB2006}, may prove insightful as well.  

\section*{Acknowledgment}
The authors thank Vivek K Goyal, Sanjoy K. Mitter, and Alan S. Willsky, as well as the anonymous reviewers for valuable comments that led to improvement of the paper.

\bibliographystyle{IEEEtran}
\bibliography{IEEEabrv,mmbre}

\end{document}